\documentclass[aps,prd,twocolumn,nofootinbib,preprintnumbers,superscriptaddresss,showpacs,amsmath,amssymb]{revtex4-1}

\usepackage{graphicx}
\usepackage{amsmath,amssymb}
\usepackage{amsfonts}
\usepackage{xspace} 
\usepackage[usenames]{color}
\usepackage{dcolumn}
\usepackage{bm}
\usepackage{mathrsfs}
\usepackage[colorlinks=true]{hyperref}
\usepackage[all]{hypcap} 
\usepackage[utf8]{inputenc} 
\usepackage{slashed}

\begin{document}
%
%

\title{Bosonic Dark Matter in Neutron Stars and its Effect on Gravitational Wave Signal}

\author{Davood Rafiei Karkevandi$^{2}$} \email{davood.rafiei64@gmail.com} 
\author{Soroush Shakeri$^{1,2}$} \email{s.shakeri@iut.ac.ir}  
\author{Violetta Sagun $^{3}$} \email{violetta.sagun@uc.pt}
\author{Oleksii Ivanytskyi$^{4}$} \email{oleksii.ivanytskyi@uwr.edu.pl} 

\affiliation{$^1$ Department of Physics, Isfahan University of Technology, Isfahan 84156-83111, Iran}
\affiliation{$^2$ ICRANet-Isfahan, Isfahan University of Technology, Isfahan 84156-83111, Iran}
\affiliation{$^3$ CFisUC, Department of Physics, University of Coimbra, 3004-516 Coimbra, Portugal}
\affiliation{$^4$ Institute of Theoretical Physics, University of Wroclaw, 50-204 Wroclaw, Poland}

\date{\today}
 \begin{abstract}
We study an impact of self-interacting bosonic dark matter (DM) on various observable properties of neutron stars (NSs). The analysis is performed for asymmetric DM with masses from few MeV to GeV, the self-coupling constant of order $\mathcal{O}(1)$ and various DM fractions. Allowing a mixture between DM and baryonic matter, the formation of a dense DM core or an extended dark halo has been explored.  We find that both distribution regimes crucially depend on the mass and fraction of  DM for sub-GeV boson masses in the strong coupling regime. From the combined analysis of the mass-radius relation and the tidal deformability of compact stars including bosonic DM, we set a stringent constraint on  DM fraction. We conclude that observations of 2$M_{\odot}$ NSs together with $\Lambda_{1.4}\leq580$ constraint, set by LIGO/Virgo Collaboration, favor sub-GeV DM particles with low fractions below $\sim 5 \%$.

\end{abstract}

\maketitle

\section{Introduction}

Despite enormous efforts during the past three decades, the nature of dark matter (DM) still remains
unknown. Currently there are several motivated particle candidates for DM such as Weakly Interacting Massive Particles (WIMPs), axions, sterile neutrinos, etc \cite{Bertone:2018krk,Bertone:2016nfn}. Research into the detectability of these particles has revealed a vast number of promising experimental facilities ranging from high-precision table-top experiments to incorporating astronomical surveys and Gravitational-Wave (GW) observations. 

The terrestrial experiments have been operating during the years in order to detect DM particles, including the recoil experiments that aim to measure the direct evi\-dence of the interaction between the new particle and nuclear or atomic targets \cite{2016PhR...627....1M,Billard:2021uyg,Roberts:2019chv,Catena:2019gfa}.  Due to the null results achieved in the direct DM searches with masses larger than a few GeV in the past years,  recently an increase of detection efforts have been conducted in the sub-GeV region, which has a mass between 1 keV and the mass of proton \cite{PhysRevLett.118.031803,PhysRevResearch.1.033105,XENON:2019gfn,Essig:2017kqs}. The sub-GeV mass range is relatively unexplored because of the experimental challenges  of detecting such light DM particles with traditional techniques. Recently, the XENON1T collaboration observed a $3.5\sigma$ excess of events from recoil electrons \cite{XENON:2020rca} which might be the evidence for the existence of DM particles with masses around $90\ \text{keV}$ \cite{Shakeri:2020wvk}. 

Other existing  DM search strategies are the decay of known particle into the DM particles, e.g, the anomalous decay of B-meson reported by the LHCb collaboration \cite{2019EPJC...79..517C}, and indirect searches which are looking for  self-annihilation signal generated by DM particles at the Galactic center \cite{2019arXiv191209486A,Slatyer:2015kla,Brdar:2017wgy}.

On the other hand, compact astrophysical objects such as neutron stars (NSs) can capture a sizable amount of DM which provide a unique astrophysical laboratory to  indirectly probe the  nature of DM. The presence of DM can substantially alter the star's structure and its thermodynamic properties leading to observable signals from astrophysical measurements \cite{Panotopoulos_2017,Nelson_2019,Ellis:2018bkr, PhysRevD.102.063028,Rezaei:2016zje, deLavallaz:2010wp,Gresham:2018rqo}. An accretion of self-annihilating DM into a NS  can be detected via increasing the luminosity and the effective temperature~\cite{Kouvaris_2008, Bhat:2019tnz,Fuller:2014rza,Acevedo:2020avd} or by modifying the  cooling curves of the star of a certain mass \cite{2019PhRvD..99d3011S, 2016PhRvD..93f5044S}. However,  DM particles with negligible annihilation rate will  settle within the NSs. This scenario is realized as  Asymmetric Dark Matter (ADM) in which a particle-antiparticle asymmetry in the dark sector exists similar to one of the baryon asymmetry in the Universe \cite{Kaplan:2009ag,Shelton:2010ta,Petraki:2013wwa,Kouvaris:2015rea}. Non-annihilating ADM whether fermionic or bosonic nature might form stable compact object named {\it{dark stars}} \cite{Narain:2006kx,Kouvaris:2015rea,Eby:2015hsq}.  Accumulation of ADM in NS could continue until  a black hole (BH) formation which can tightly constraint ADM models by the presence of old NSs \cite{McDermott:2011jp,Bertone:2007ae,Kouvaris:2010jy,Kouvaris:2011fi,Acevedo:2020gro,Khlopov:1985jw}.

 It has been shown that an accretion of massive ADM particles  can significantly reduce the maximum mass of the host NS making the two solar mass limit unattainable \cite{PhysRevC.89.025803, Leung:2011zz, Li:2012ii, Deliyergiyev_2019}. Thus, observational fact of an existence of two heavy pulsars, i.e, PSR J0348+0432 with mass $2.01\pm 0.04M_{\odot}$ \cite{Antoniadis:2013pzd} and PSR J0740+6620 of $2.14^{+0.10}_{-0.09}M_{\odot}$   \cite{NANOGrav:2019jur}, enables us to constrain the properties of massive DM particles and their fraction inside the compact stars. On the other hand, light DM particles lead to formation of an extended halo around the NS and could increase its total gravitational mass~\cite{PhysRevD.102.063028, Nelson_2019}. 

Based on the NS observations, it has been argued that the presence of light bosonic ADM without self-interaction has been excluded in the mass range $2\ \text{keV}$ - $16\ \text{GeV}$ due to the BH formation \cite{Kouvaris:2011fi}. Furthermore,  a similar result has been reported about tight constraints  on the noninteracting scalar ADM with masses between 5 MeV and 13 GeV \cite{McDermott:2011jp}. 
The observed discrepancy in the low-mass region reported in Ref. \cite{Kouvaris:2011fi} and Ref. \cite{McDermott:2011jp} is because of an inclusion of an effect of neutron degeneracy on the capture rate in the latter case. However, it was shown that taking into account a repulsive self-interaction among DM particles prevents instability issues related to the BH formation \cite{Bell_2013,Kouvaris:2013awa,Nelson_2019,Mielke:2000mh,Ho:1999hs}. In addition,  self-interacting DM can  resolve  a series of issues in  the collisionless cold DM (CCDM) scenario for the small-scale cosmological observations  \cite{Rocha:2012jg,Vogelsberger_2012,2013MNRAS.431L..20Z,Arguelles:2015baa}. The self-interaction cross section per unit DM mass in range $0.1\ \text{cm}^2/\text{gr}\lesssim\sigma/m\lesssim 1\ \text{cm}^2/\text{gr} $ is sufficient  to explain various inconsistencies  between numerical simu\-lations and observational results in CCDM paradigm 
\cite{Giudice:2016zpa,Eby:2015hsq,AmaroSeoane:2010qx,Peter:2012jh,Kaplinghat:2015aga}. Moreover, it has been shown that a  self-interacting  complex scalar field can form a Bose-Einstein condensate (BEC) which yields an attractive  solution for the DM Galactic halo and other astronomical DM issues  \cite{PhysRevD.68.023511,Suarez:2013iw,Suarez:2015fga,Boehmer:2007um,Li:2013nal,Chavanis:2016ial,Sikivie:2009qn,Harko:2011xw,Chavanis:2011cz,Suarez:2016eez}.

The stability of a self-gravitating system of fermionic DM particles without self-interaction is provided by the Fermi pressure. For bosonic DM the only source of pressure against the gravitational contraction comes from the uncertainty principle leading to  formation of Boson Stars (BSs), for a comprehensive review on BSs see \cite{Schunck:2003kk,Liebling:2012fv,Visinelli:2021uve}. Historically, the idea of BS  was  proposed by \citet{Kaup:1968zz}, \citet{Ruffini:1969qy}, they showed that  BSs consisting of noninteracting particles have much lower maximum mass compared to their fermionic counterparts. However,  introducing the repulsive interaction between bosons, e.g, proposed by \citet{Colpi:1986ye}, drastically changes the physical properties of BSs. Depending on the mass and strength of the self-interaction between DM particles,  BSs of stellar mass could have observable signatures at GW detectors \cite{Pacilio:2020jza,Giudice:2016zpa}. Therefore, either self-interacting bosonic ADM as the complex scalar field or ultra light axions without self-interaction could form a dark BS of the stellar mass \cite{Gleiser:1988rq,Kusmartsev:1990cr,Kolb:1993zz,Schunck:1996he,Mielke:2000mh,Chavanis:2011cz,Maselli:2017vfi,Eby:2015hsq,2011PhRvD..84d3532C,2011PhRvD..84d3531C,Visinelli:2017ooc,Chavanis:2017loo,Kouvaris:2019nzd}. As other possibilities, the dark BS can be formed in terms of a Bose-Einstein gravitational condensation described by  Gross-Pitaevskii-Poisson 
equation  \cite{Chavanis:2011cz,Li:2012sg}, or it can be made of bosons with a repulsive self-interaction described within the mean-field approximation and general relativity \cite{Agnihotri:2008zf}.  Both above mentioned dark BS models were utilized as a DM component within NSs \cite{Nelson_2019,Li_2012,Ellis:2018bkr}. 

Generally, three different scenarios can be realized for a DM admixed NS \cite{Mukhopadhyay:2016dsg, Li_2012,Goldman_2013, Tolos_2015}  :\\
{\it(i)} DM is condensed in a NS core. In this case, radius of DM component ($R_{D}$) is smaller than radius of baryonic matter (BM) component ($R_{B}$), i.e, $R_{B}>R_{D}$;\\
{\it(ii)} DM distributed in entire NS with $R_{B}=R_{D}$; \\ 
{\it(iii)} DM creates an extended halo around a NS with $R_{D}>R_{B}$.\\
The above mentioned configurations are valid for only gravitational interaction between BM and DM,  allowing a mixture of both  fluids at the core  and the possible appearance of one of them at the outer shell of the combined object \cite{Nelson_2019,Ellis:2018bkr,PhysRevC.89.025803, Goldman_2013, Tolos_2015, Sandin_2009, Ciarcelluti:2010ji, Mukhopadhyay:2016dsg, Deliyergiyev_2019, Mukhopadhyay:2015xhs, PhysRevD.102.063028}.  

In the case of sufficiently strong nongravitational interaction between both components their mixing is prevented leading to two more scenarios. One of them corresponds to a star with a pure baryonic core surrounded by a DM halo, while the other one describes a star with a pure DM core and a baryonic shell \cite{Gresham:2018rqo,Li_2012, Zhang:2020pfh,Zhang:2020dfi}. Meanwhile, in the presence of the nongravitational interactions between BM and DM, the whole  system can be described by a single equation of state (EoS) obtained from the relativistic mean-field model \cite{Panotopoulos_2017,Gresham:2018rqo,Das:2018frc,Das:2020vng,Sen:2021wev,Das:2021wku}.

The recent detections of GWs from  the binary NS mergers  opened a new window for probing DM particles \cite{Abbott_2017,Abbott_2020}. The impact of ADM particles on the internal structure of NSs can be considered through GW signals especially during the post-merger stage \cite{Ellis:2017jgp, Bezares:2019jcb, Bezares:2018qwa, Horowitz:2019aim, Bauswein:2020kor}. The  GW signal is
also sensitive to  the deformation effects  of binary NSs
 during the inspiral  phase. This information encodes in  the tidal deformability parameter which gives valuable clues into the  EoS of NSs \cite{Hinderer:2007mb,Hinderer:2009ca,Postnikov:2010yn,Zhao:2018nyf,Han:2018mtj}. The upper bound on tidal deformability $\Lambda\leq800$ for $M=1.4M_{\odot}$ was obtained by the GW observation at the $90\%$ confidence level for GW170817 created by coalescence of the binary NSs \cite{Abbott_2017}. An improved estimate of $\Lambda_{1.4}=190^{+390}_{-120}$ has been reported in \cite{Abbott:2018exr}. It is worth mentioning that most of BM EoSs obtained from many-body theories, taking into account the realistic nucleon-nucleon interactions \cite{Nelson_2019,Gandolfi:2011xu,Hebeler:2015hla,Tews:2018kmu,Hebeler:2010jx,Goudarzi:2016uos,Fattoyev:2017jql,Dengler:2021qcq}, produce the tidal deformability in the range of 100-500. The presence of DM in NSs will alter the tidal deformability which  can be utilized to probe the parameter space of DM model and the amount of DM within NSs to be consistent with observational constraints
  \cite{Nelson_2019, Ellis:2018bkr, Quddus:2019ghy, Ciancarella:2020msu, Zhang:2020dfi, LeTiec:2020spy, Das_2019, Sen:2021wev}. Meanwhile, multimessenger observations of NSs from combining GW detections (by e.g, LIGO/Virgo/KAGRA \cite{LIGOScientific:2021qlt}) with x-ray (by e.g, NICER \cite{2019ApJ...887L..24M, Raaijmakers_2020}) and radio (by e.g, SKA \cite{2015aska.confE..43W}) data can be used to examine the presence of DM inside or around the NSs \cite{PhysRevLett.126.181101}.

In this work, we study an effect of  
bosonic ADM on the compact star properties such as the maximum mass and  tidal deformability for different DM distribution regimes.  We  focus here on DM mass ranging from a few MeV to GeV and the strong coupling regime with a coupling constant of order $\mathcal{O}(1)$. The DM component is treated as a self-repulsive
complex scalar field described by the EoS proposed by \citet{Colpi:1986ye}. Further on we will refer to it as a bosonic self-interacting DM (SIDM). This EoS is demonstrated to produce BSs of high enough mass to be consistent with typical NSs described within hadronic EoSs \cite{Maselli:2017vfi}. On the other hand, the baryonic component is modeled by the unified EoS with induced surface tension (IST) that was successfully applied to describe the nuclear matter, heavy-ion collision data and dense matter existing inside NS \cite{2018NuPhA.970..133B, Sagun:2018cpi, Sagun11:2018sps}.

We show that depending on the mass, fraction and the strength of DM self-interaction, a DM halo or DM core can be formed which will modify the GW emission during the coalescence of two compact stars. Taking into account two key observable constraints of NSs, i.e,   maximum  mass $M_{T_{max}}\geq2M_{\odot}$  and  tidal deformability $\Lambda_{1.4}\leq580$, we set an upper limit on the fraction of sub-GeV bosonic DM inside NSs which disfavors the values above $\sim5\%$. The considered fractions include the conservative values obtained by the DM accretion from the surrounding medium \cite{Deliyergiyev_2019,DelPopolo:2019nng, Baryakhtar:2017dbj,Bramante:2015dfa}, as well as higher values based on possible scenarios of its augmentation, e.g, enhanced production of DM during the supernova explosion stage \cite{Nelson_2019}, absorption of primordial DM clumps \cite{Ellis:2018bkr,Goldman_2013}, etc.

The paper is organized as follows. In Sec. \ref{secbosonicdm} we describe the EoSs for bosonic DM and BM. In Sec. \ref{secdistribuition} we show the distribution of DM for different values of coupling constant, mass of DM particles and their fraction. Sections \ref{secmass} and \ref{sectides} are devoted to the analysis of an effect of DM on maximum mass and tidal deformability of NSs. In Sec. \ref{secfrac} we present a constraint on the mass of DM particles and their fraction.  Finally we briefly discuss different scenarios for the presence of DM inside compact stars in Sec. \ref{accretion}. The results are summarized in Sec. \ref{secconclusion}. We use units in which $\hbar=c=G=1$.

\section{Dark and Baryon Matter models}
\label{secbosonicdm}

\subsection{Dark matter equation of state }

In the following, we treat DM as massive self-interacting bosons carrying conserved charge. Such particles are described by a complex scalar field with the self-interaction potential $V(\phi)=\frac{\lambda}{4}|\phi|^4$, where $\lambda$ is a dimensionless coupling constant \cite{Colpi:1986ye,Maselli:2017vfi}. In this setup a coherent scalar field is governed by Klein-Gordon equation and can potentially form Bose Einstein condensate (BEC) if the temperature is sufficiently low \cite{PhysRevD.68.023511,Chavanis:2011cz,Suarez:2013iw}. In this work, we assume DM to exist at zero temperature, thus, leading to its total condensation. Thermal fluctuations are also suppressed in this case. This justifies treating BEC of DM within the mean-field approximation. The corresponding EoS of bosonic matter with repulsive self-interaction is given by
\begin{eqnarray}\label{e1}
P&=&\frac{m_{\chi}^{4}}{9\lambda}\left( \sqrt{1+\frac{3\lambda}{m_{\chi}^{4}}\rho}-1\right)^{2}\, , 
\end{eqnarray}
where $m_\chi$ is the DM particle mass. Derivation of Eq. (\ref{e1}) is given in Appendix. This EoS is obtained in locally flat space-time, which requires small gradients of metrics and absence of the anisotropy issues \cite{Chavanis:2011cz,Mielke:2000mh,Schunck:2003kk,AmaroSeoane:2010qx,Gleiser:1988rq,Kusmartsev:1990cr,Suarez:2016eez,Chavanis:2021tva}. This condition is provided at 
\begin{eqnarray}
\lambda \gg 4\pi (m_{\chi}/M_{Pl})^{2}=8.43\times 10^{-36} \left( \frac{m_{\chi}}{100\,\text{MeV}} \right)^{2} \,
\end{eqnarray}
which is well inside the range considered in this work. EoS (\ref{e1}) can be approximated by a polytropic equation, its corresponding index changes from 2 to 1 at low and high densities, respectively. The critical density of switching between the two regimes is estimated as
\begin{eqnarray}\label{d}
\rho_{c}=\frac{m_{\chi}^{4}}{3\lambda}= \frac{4.3}{\lambda}\,  \left(\frac{m_{\chi}}{100 \,\text{MeV}} \right)^{4} \text{MeV}/\text{fm}^{3}.
\end{eqnarray}

The EoS (\ref{e1}) was applied to study hypothetical compact objects composed of bosonic DM, i.e, BSs \cite{Colpi:1986ye,Maselli:2017vfi,Chavanis:2011cz}. The maximum mass of such objects was found to be
\begin{eqnarray}\label{emax}
M_{\text{max}}^{\text{BS}}\approx 0.06 \lambda^{1/2} M_{\text{Ch}} \approx 10\, M_{\odot} \lambda^{1/2}  \left( \frac{100\,\text{MeV}}{m_{\chi}} \right)^{2} \,
\end{eqnarray}
where $M_{\text{Ch}}\approx M_{Pl}^{3}/m_{\chi}^{2}$ is the Chandrasekhar mass. According to Eq. (\ref{emax}), stellar mass BSs can be formed for $\lambda\sim\mathcal{O}(1)$ and $m_{\chi}\sim\mathcal{O}$(100 MeV)   \cite{Schunck:2003kk,Pacilio:2020jza}.  On the other hand, the maximum compactness $\mathcal{C}_{(max)}=M/R$ of  the BS configurations corresponding to our desired parameter space is about 0.16   which  is far  below the BH formation limit \cite{AmaroSeoane:2010qx}.  Lower DM mases and/or higher couplings lead to stiffening of the resulting EoS. It has been further proven that solutions of the Tolman-Oppenheimer-Volkof (TOV) equations \cite{Tolman:1939jz,Oppenheimer:1939ne} with the present EoS of DM are  self-similar \cite{Maselli:2017vfi}, allowing general statements about TOV solutions without scanning over the whole parameter space.

\subsection{Baryonic matter equation of state}
\label{secist}
To model BM we utilize the IST EoS  developed in Refs. \cite{2014NuPhA.924...24S,Sagun:2018cpi}. It reproduces four first virial coefficients of the gas of hard spheres providing an accurate account of the short range particle repulsion of the hard-core type \cite{2014NuPhA.924...24S}. The long-range attraction between baryons is incorporated to the present model within the mean field framework \cite{Sagun:2018cpi}. This EoS is fitted to the nuclear matter ground state properties \cite{2014NuPhA.924...24S}, fulfills the proton flow constraint \cite{Ivanytskyi:2017pkt}, reproduces multiplicities of hadrons measured in heavy ion collisions \cite{Sagun:2017eye,2018NuPhA.970..133B,2018NuPhA.970..133B}. Supplemented by the conditions of electric neutrality and $\beta$-equilibrium the IST EoS was successfully applied to modeling NS \cite{ Sagun:2018cpi, Sagun11:2018sps}.

In the present study we utilize the model set up proposed in Ref. \cite{Sagun_2020} (see set B). It yields the nuclear asymmetry energy and its slope at saturation density $J = 30.0$ MeV and $L=93.19$ MeV, respectively, maximum NS mass $M_{max}=2.08 M_{\odot}$ and radius of the $1.4 M_{\odot}$ star equals to $R_{1.4}=11.37$ km.

Following Ref. \cite{PhysRevD.102.063028} the NS crust is modeled by the polytropic EoS with $\gamma=4/3$ that mimics the atomic structure of the outer and inner crusts. The crust EoS was smoothly matched to the IST EoS at density 0.09 $fm^{-3}$.

\section{Dark Halo and Dark Core Formation Regimes}
\label{secdistribuition}

In order to study the compact objects formed by the admixture of BM and bosonic SIDM, we consider two-fluid TOV formalism where each component is described as a perfect fluid. Due to the negligibly weak interaction between DM and BM we consider their interaction only through gravity \cite{Sandin_2009,Ciarcelluti:2010ji,Deliyergiyev_2019,Tolos_2015,Ellis:2018bkr}. In this case, the energy-momentum tensors of each component are conserved separately
(for an explicit derivation see Refs. \cite{PhysRevD.102.063028,PhysRevC.89.025803}) and the system of equations for relativistic hydrostatic equilibrium is defined as
\begin{eqnarray}\label{e6}
\frac{dp_{\text{B}}}{dr}&=& -\left( p_{\text{B}} +\epsilon_{\text{B}} \right) \frac{d\nu}{dr}\,,\\ 
\frac{dM_{\text{B}}}{dr} &=& 4 \pi \epsilon_{\text{B}} r^2\,, \\ 
\frac{dp_{\text{D}}}{dr} &=& -\left( p_{\text{D}} +\epsilon_{\text{D}} \right) \frac{d\nu}{dr}\,, \\ 
\frac{dM_{\text{D}}}{dr} &=& 4 \pi \epsilon_{\text{D}} r^2\,, \\
\frac{d\nu}{dr}&=& \frac{(M_\text{B}+M_\text{D})+4\pi r^3(p_\text{B}+p_\text{D})}{r(r-2(M_\text{B}+M_\text{D}))}\, ,
\label{e6d}
\end{eqnarray}
where $p_{B} (p_{D})$ and $\epsilon_{B} (\epsilon_{D})$ are pressure and energy density of BM (DM) component, and $r$ is the distance from the center of a star. Thus, total pressure $p=p_{B}+p_{D}$ and energy density $\epsilon=\epsilon_{B}+\epsilon_{D}$ have two contributions from BM and DM. 

The system of Equations (\ref{e6}-\ref{e6d}) was obtained from the Einstein ones for spherically symmetric metric
\begin{eqnarray}
ds^{2}=-e^{2\nu(r)}dt^{2}+e^{2\lambda(r)}dr^{2}+r^{2}d\Omega^{2},
\end{eqnarray}
where is  $\lambda$ and $\nu$ are the metric functions.

In general to solve single fluid TOV equations \cite{Tolman:1939jz,Oppenheimer:1939ne} the boundary conditions should be determined. In the case of the two-fluid formalism, two sets of boundary conditions for DM and BM  have to be considered \cite{Sandin_2009,Ciarcelluti:2010ji}. For the fixed values of central pressure, $p_{B}$ and $p_{D}$, and mass in the center of a star, $M_{B}(r\simeq0) = M_{D}(r\simeq0) \simeq 0$, we  performed a numerical integration of Eqs. (\ref{e6}-\ref{e6d}) up to a radius at which  pressure of one of  the components vanishes. In principle this radius can be realized as DM core radius $R_{D}$ or BM core radius $R_{B}$. 

Thus, for i) and ii) scenarios described in the Introduction for which DM is distributed inside the NS ($R_{B}\geq R_{D}$) we continue the numerical integration to reach the visible radius of the star where $p_B(R_B)=0$. In this case the total gravitational mass of the star is defined by

\begin{eqnarray}\label{e10}
M_{T}=\int_{0}^{R_B} 4\pi r^{2} [\epsilon_B (r) + \epsilon_D (r)] dr.
\end{eqnarray}

For a NS surrounded by an extended DM halo (see iii) scenario in the Introduction), i.e, $R_{D}>R_{B}$, where $p_{B}(r>R_{B})=0$, in order to find the total mass one has to replace the upper limit of the integration in Eq. (\ref{e10}) by $R_{D}$. Consequently, the total gravitational mass of a DM admixed NS is
\begin{eqnarray}
M_{T}=M_{B} (R_{B})+M_{D}(R_{D}).
\end{eqnarray}
However, the observable radius of the star is still defined by $R_{B}$, this is due to the visibility of $R_{B}$ compared to $R_D$ and technical difficulties in direct detection of dark radius $R_{D}$.

The DM fraction is  a  crucial parameter in our analysis, which characterizes the amount of DM in a DM admixed NS and is defined as
\begin{equation}
    F_{\chi}=\frac{M_D(R_{D})}{M_T}.
\end{equation}
As the distribution of DM depends on the particle's mass $m_{\chi}$, fraction $F_{\chi}$ and the value of the coupling constant $\lambda$ we perform a thorough analysis to show the role of each parameter. Thus, Fig. \ref{E,M-R400} shows energy density and mass profiles for  $m_{\chi}=400$ MeV, $\lambda=\pi$ values and different DM fractions $F_{\chi}$ between $10\%- 50\%$. It was implemented by fixing pressure of both components in the center in such a way to obtain the desired fraction. For better understanding each matter component is depicted separately, i.e, BM (dashed lines) and DM  (solid lines). From Fig. \ref{E,M-R400}, we see that a DM core with $R_{D}\approx 5\text{km}$ is embedded in a baryonic star with bigger radius. An increase of DM fraction from $10\%$ to  $50\%$ leads to an increase of the size and mass of the DM core, while these properties of baryonic component decrease and it becomes more compact.

\begin{figure}[h]
    \centering
    \includegraphics[width=3.2in]{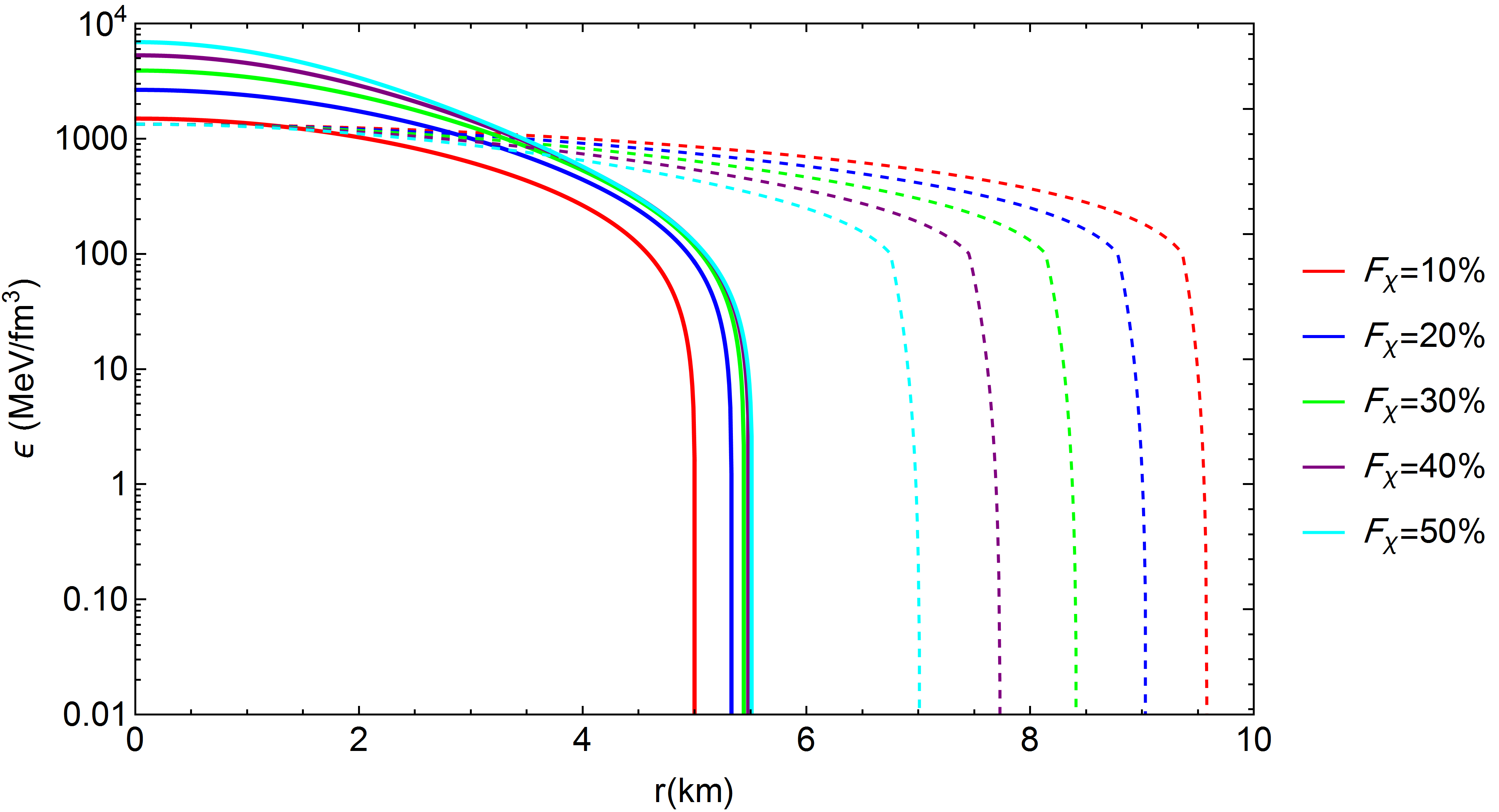}  \includegraphics[width=3.1in]{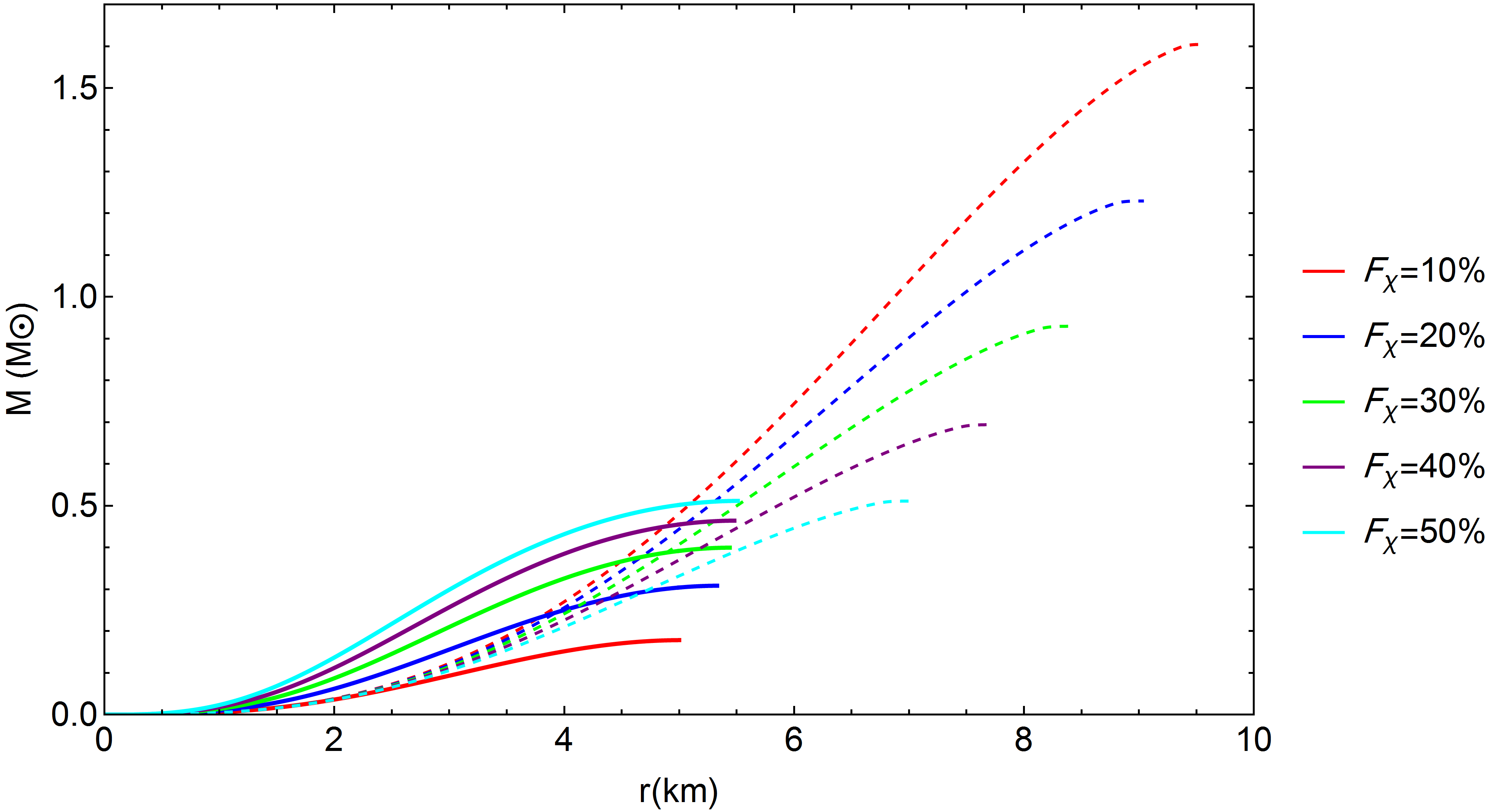}
    \caption{Energy density (upper panel) and enclosed mass (lower panel) as a function of star radius for DM admixed NSs. Calculations are made for $m_{\chi}=400$ MeV, $\lambda=\pi$ and different DM fractions between $10\%- 50\%$. Solid and dashed lines correspond to DM and BM components, respectively. For the considered values of parameters the DM core is formed inside a NS.}
    \label{E,M-R400}
\end{figure}

Another type of behavior is demonstrated in Fig. \ref{E,M-R100} where energy density and mass profiles for $m_{\chi}=100$ MeV are presented. As it is seen the light DM particles lead to a formation of a halo around BM star with much larger radius that is a function of DM fraction. Higher fractions lead to an increase of mass and radius of DM halo. 

\begin{figure}[h]
    \centering
    \includegraphics[width=3.1in]{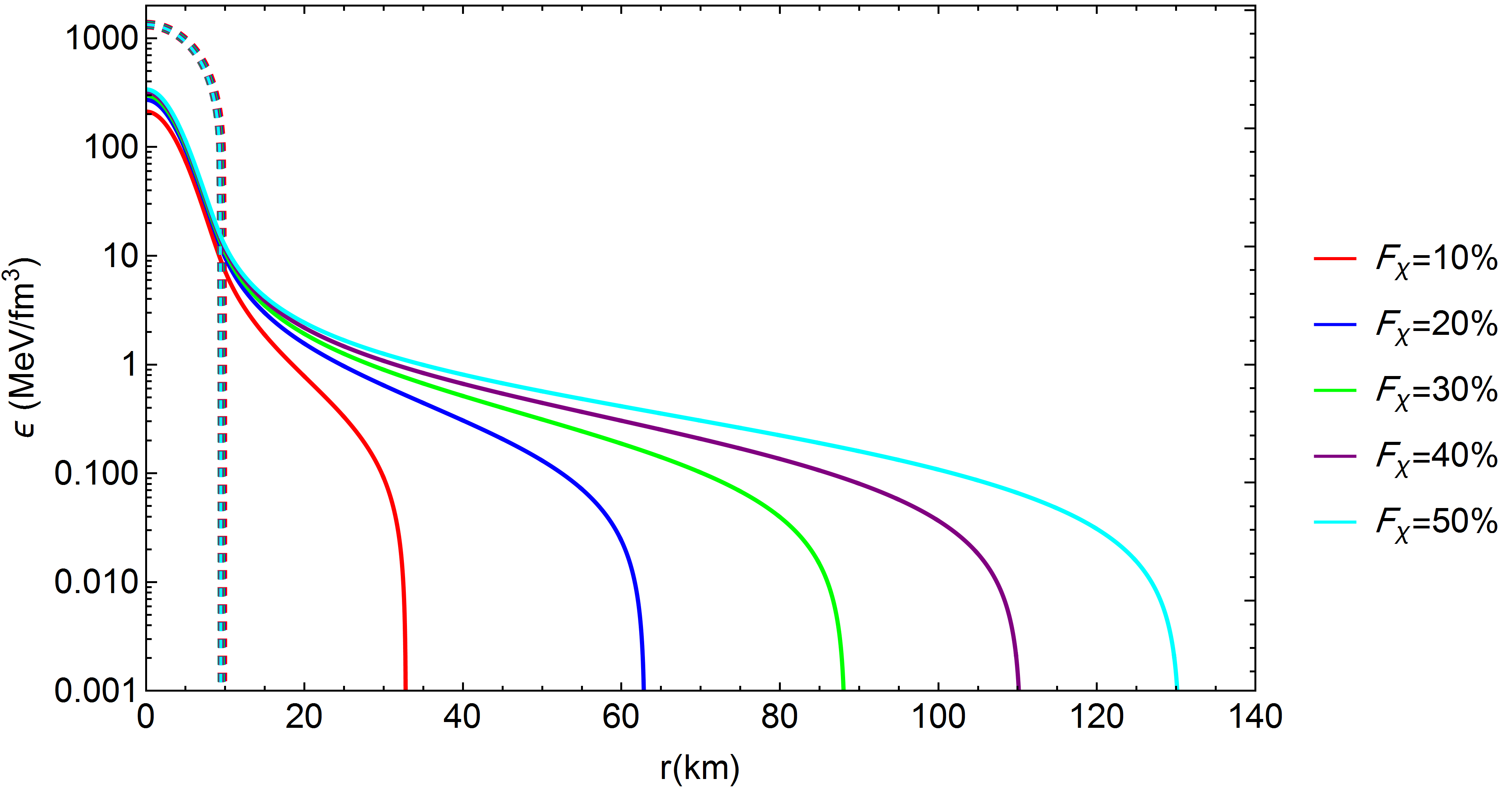} 
    \vspace{0.1cm}
    \includegraphics[width=3.1in]{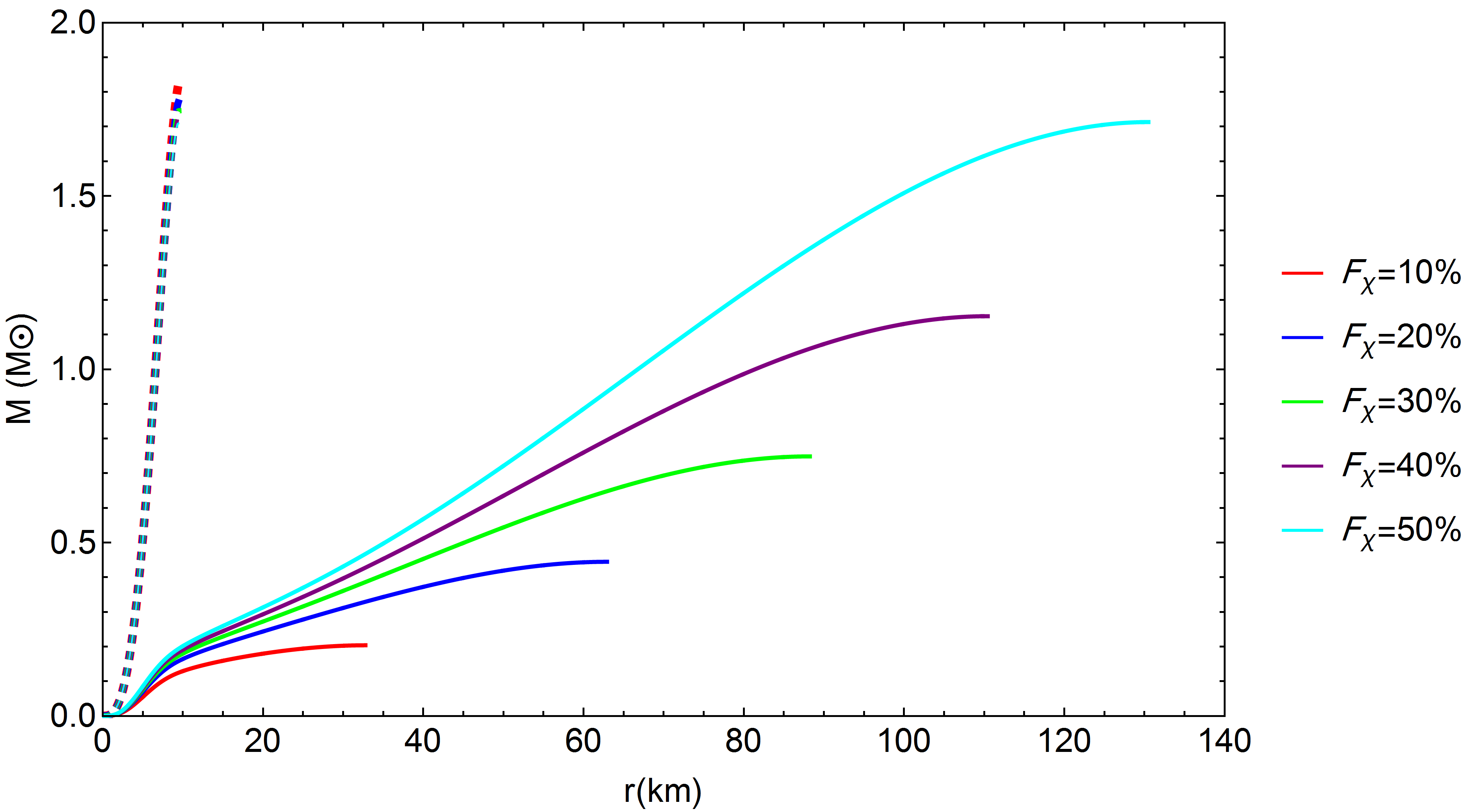}
    \caption{Energy density (upper panel) and enclosed mass (lower panel) profiles for DM admixed NSs. Calculations are made for $m_{\chi}=100$ MeV, $\lambda=\pi$ and different DM fractions between $10\%-50\%$. Solid and dashed lines correspond to DM and BM components, respectively. For the considered values of parameters the DM halo is formed around a NS.}
    \label{E,M-R100}
\end{figure}

\begin{figure}[!h]
    \centering
    \hspace{0.38in}
    \includegraphics[width=3.4in]{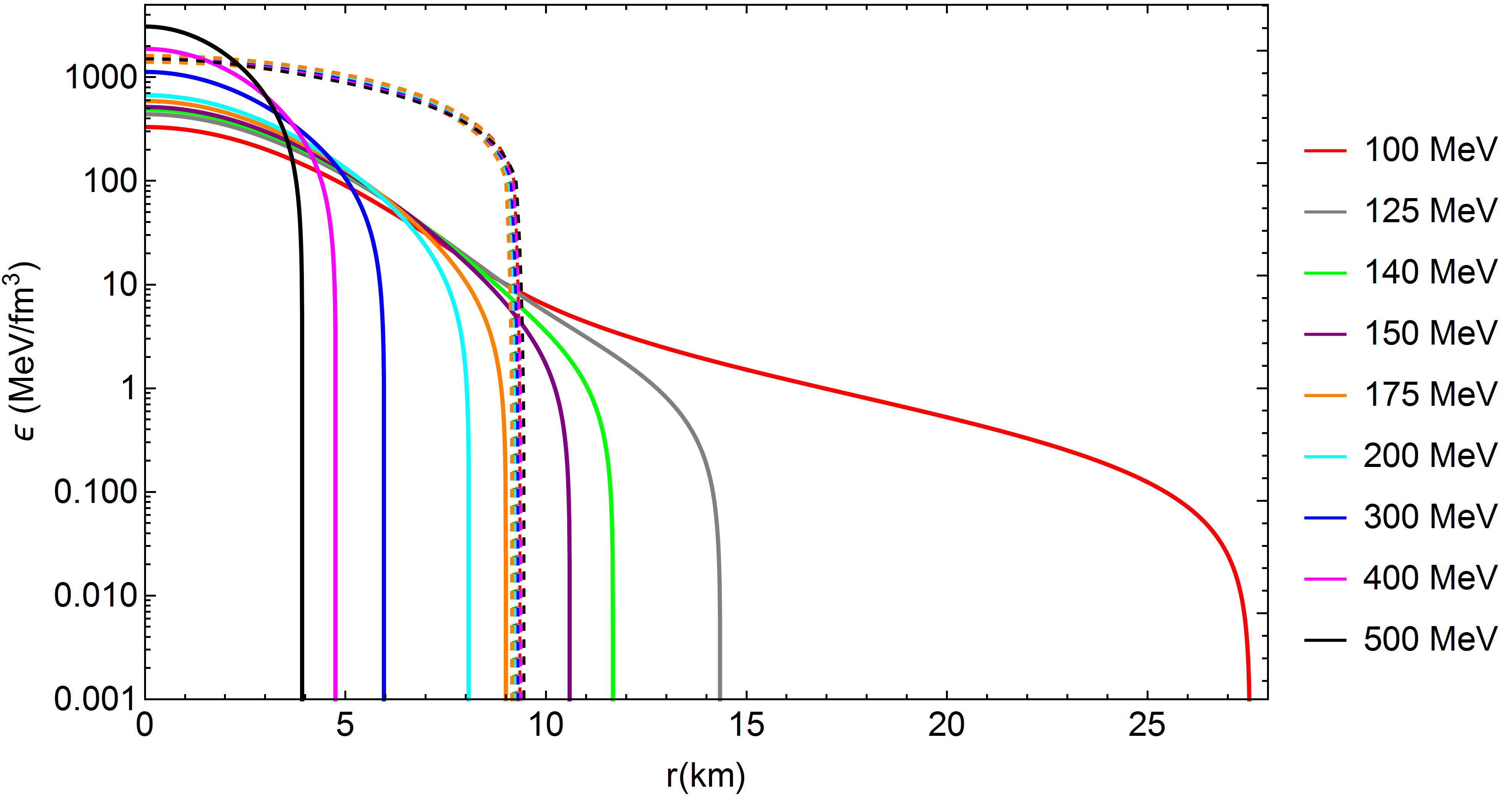}
    \label{fig:event-electron}
    \\
    \centering
    \hspace{0.45in}
    \includegraphics[width=3.4in]{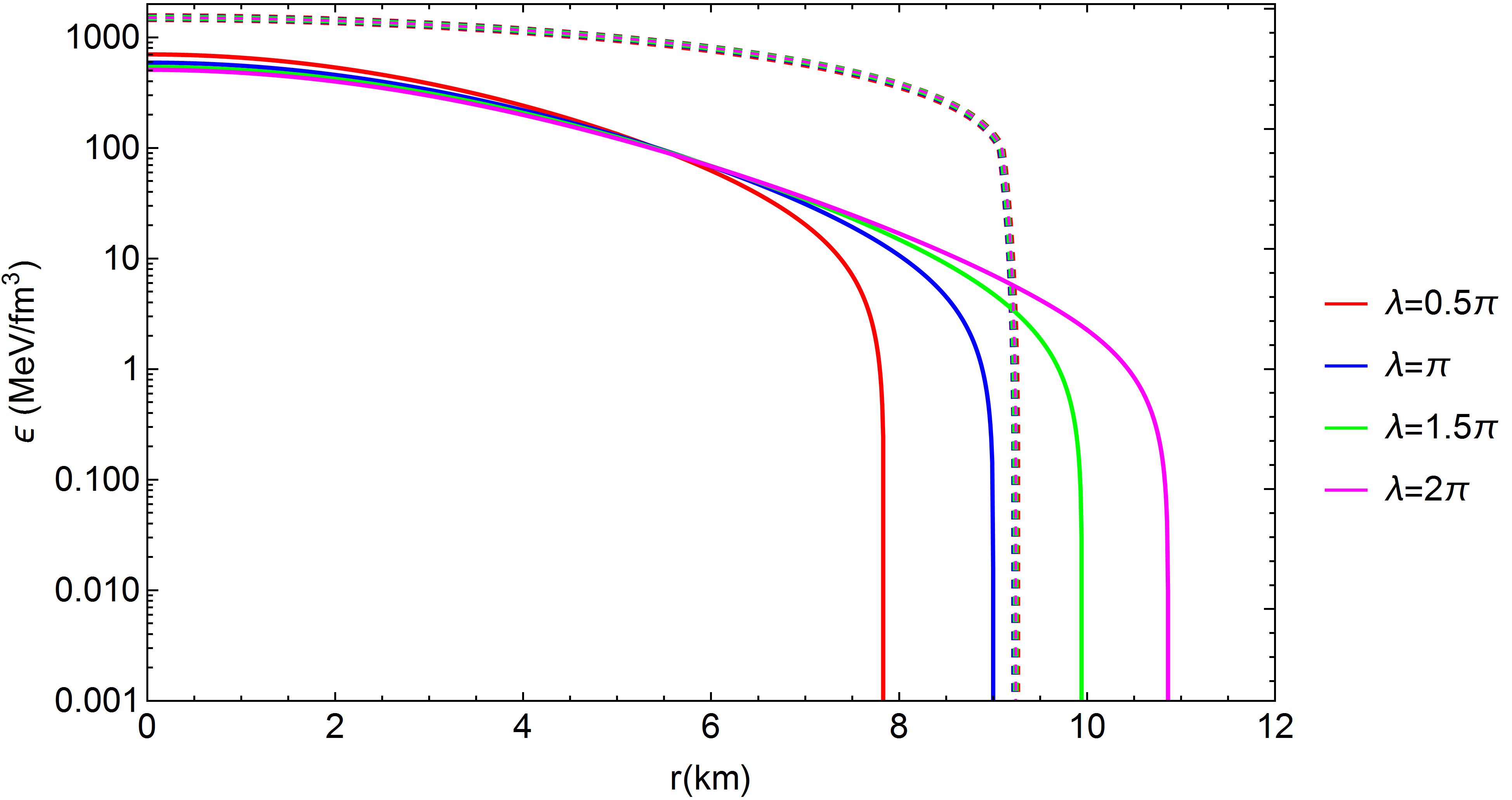} 
    \label{fig:event-electron}
    \\
     \hspace{.5in}
    \centering
      \includegraphics[width=3.4in]{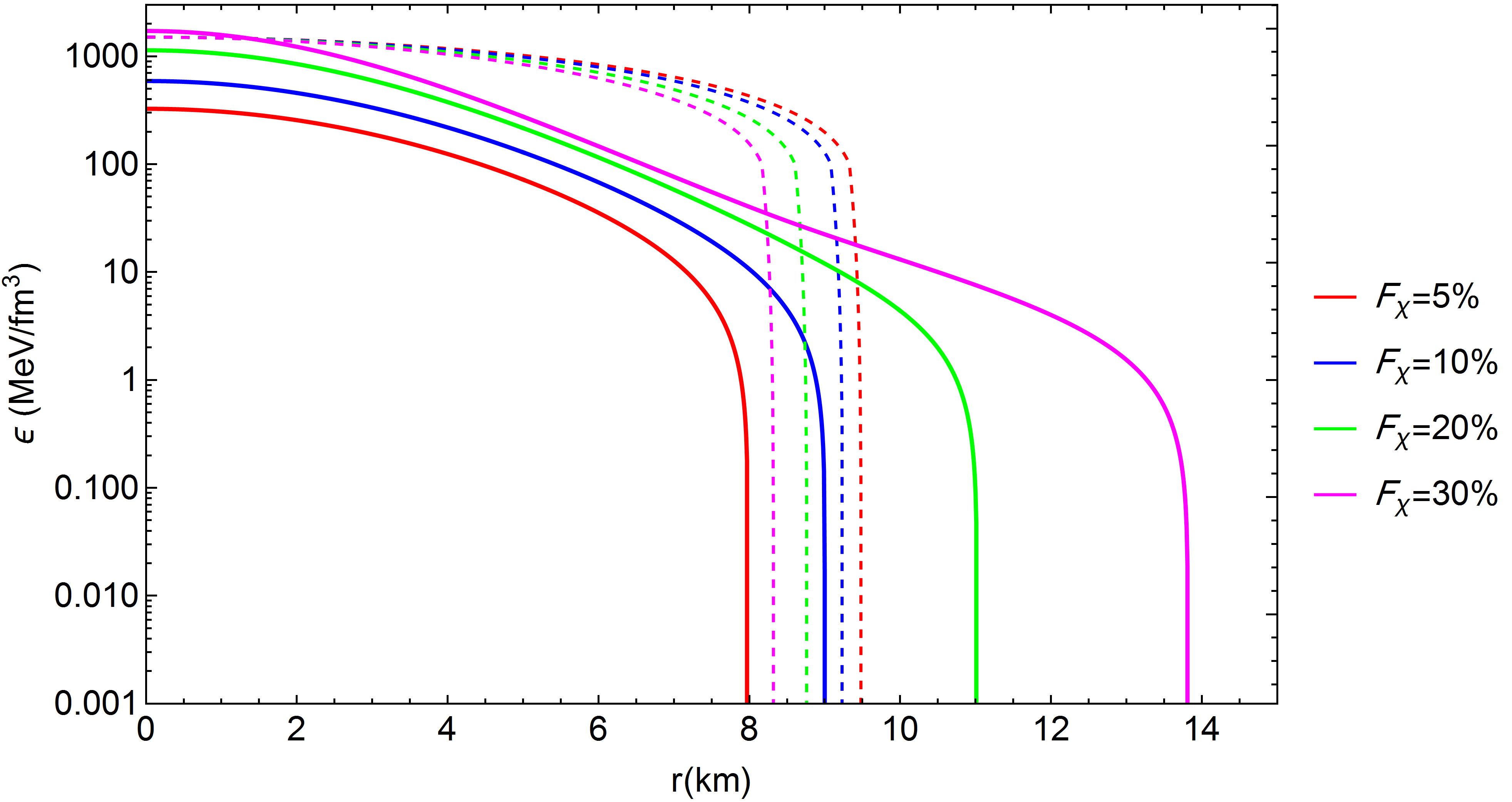}
    \caption{Energy density profiles of DM admixed NSs for different set of model parameters: $\lambda=\pi$, $F_{\chi}=10\%$ and different values of boson mass (upper panel); $m_{\chi}=175$ MeV, $F_{\chi}=10\%$ and different values of coupling constant (middle panel);  $m_{\chi}=175$ MeV, $\lambda=\pi$ and different values of DM fraction $F_{\chi}$ (lower panel).}   
    \label{core-halo}
\end{figure}

A comparison of Figs. \ref{E,M-R400} and \ref{E,M-R100} makes us to conclude that a transition from DM core to halo occurs from $m_{\chi}=400$ MeV  to $100$ MeV 
for $\lambda=\pi$ and different values of $F_{\chi}$. To spot the exact value of $m_{\chi}$ at which this transition happens,  we plot the energy density profiles for BM and DM separately with $m_{\chi}$ varies from  100 MeV to 500 MeV at fixed $\lambda$ and $F_{\chi}$. As it is seen on the upper panel of Fig. \ref{core-halo} at $m_{\chi}\approx 175$ MeV radii of both components coincide ($R_{B}\approx R_{D}$), while a slight decrease of $m_{\chi}$ leads to the formation of halo structure with  $R_{D}>R_{B}$. In the opposite case, the DM core will be formed for more massive SIDM particles. The middle and lower panels  show how the DM distribution is changed  by varying the value of coupling constant and DM fraction. 
By a thorough analysis of an effect of model parameters from Fig. \ref{core-halo}, one can show that a DM halo is always formed around NS for $F_{\chi}=10\%$ and $\lambda$ in the range between $0.5\pi-2\pi$ for $m_{\chi}\leq140$ MeV. A compatible result has been obtained recently in Ref. \cite{PhysRevD.102.063028} but for fermionic DM without any self-interaction. 

Therefore, as a general behavior we can conclude that light DM particles with $m_{\chi}<200$ MeV tend to form halo around a NS, while heavier ones for low DM fractions would mainly create a DM core inside a compact star. However, for massive DM particles it would be still possible to form a DM halo for high values of $F_{\chi}$. More detailed consideration of the role of DM fraction in the formation of DM core and DM halo will be presented in the following section. 

\section{Mass-Radius Relation in the Presence of Bosonic DM}
\label{secmass}

The mass-radius (M-R) relations for DM admixed NSs are shown in Figs. \ref{M-R,mass,lambda} and \ref{M-R,fraction} in which $M=M_{T}=M_{B}+M_{D}$. Here $R$ is the outermost radius of the star which is determined by  $R_{B}$ for (i) and (ii) scenarios and by $R_{D}$ for iii) scenario that includes a DM halo formation. A solid black curve on each panel corresponds to the M-R relation  for pure baryonic stars described by the IST EoS. Gray dashed horizontal line indicates the $2M_{\odot}$ maximum mass limit for NS, magenta and cyan regions mark causality and GR limits, respectively.

  \begin{figure}[!h]
    \centering
\hspace*{-0.1cm}\includegraphics[width=3.0in]{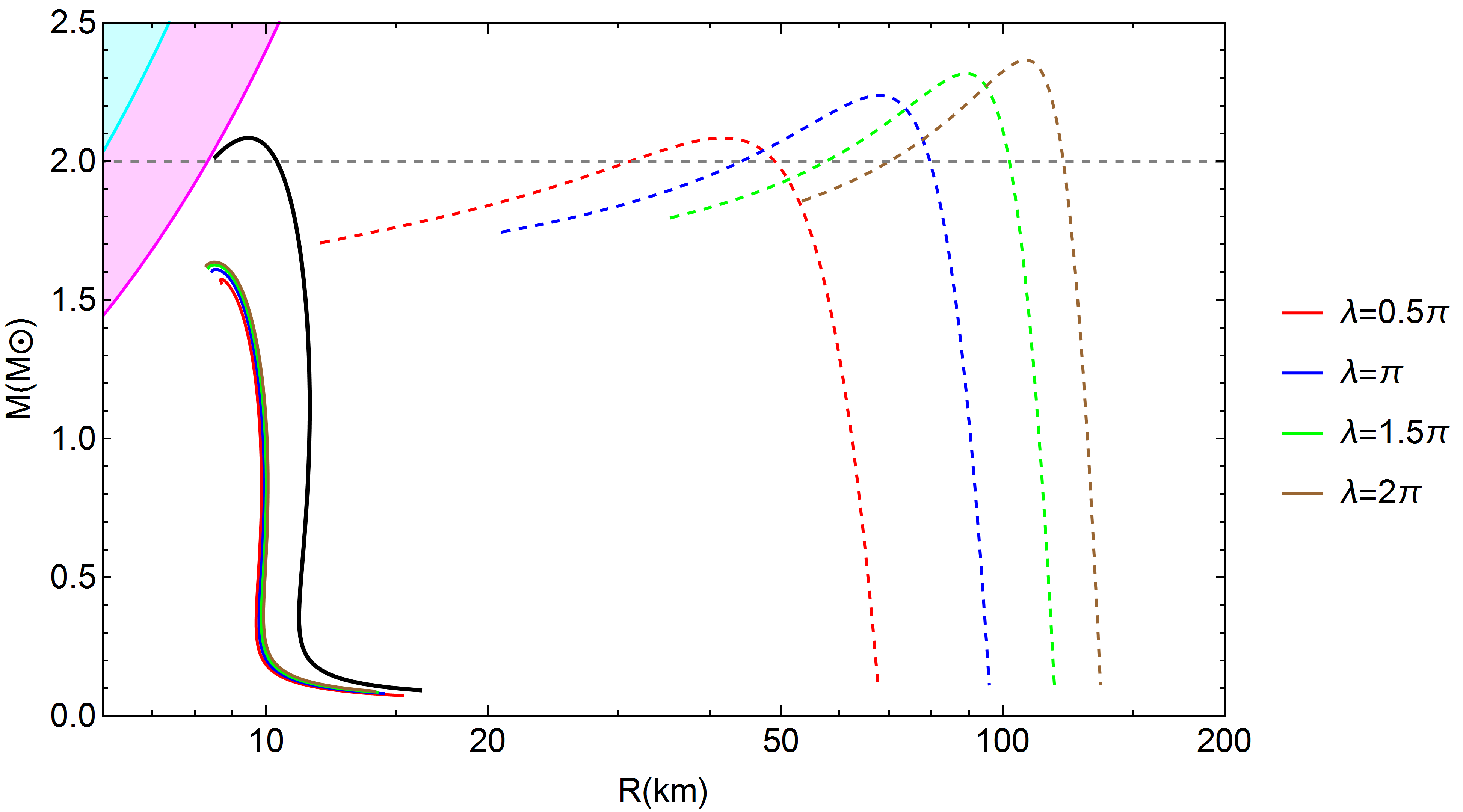}  
    \includegraphics[width=3.1in]{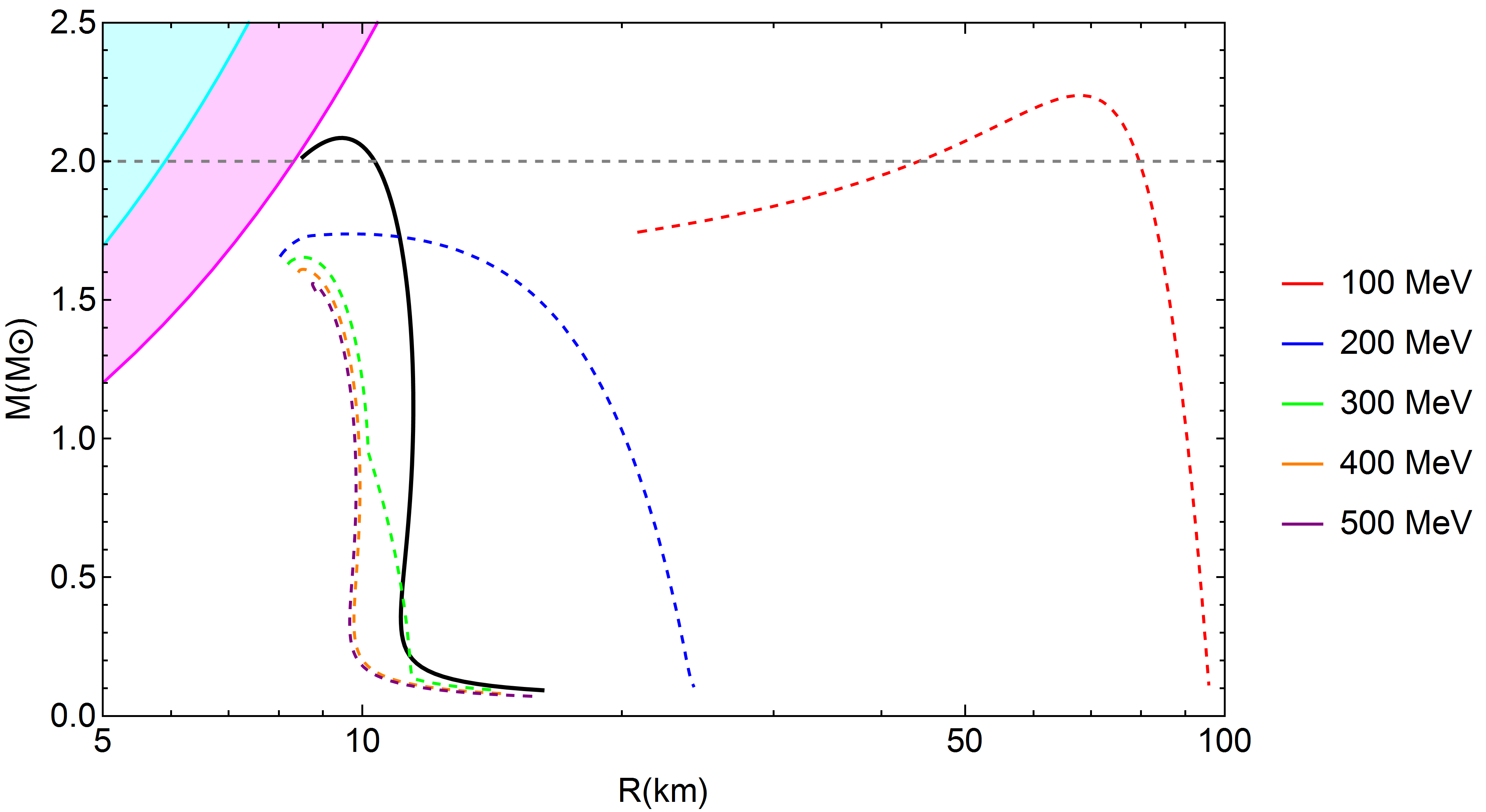}
    \caption{M-R relations of DM admixed NSs with $F_{\chi}=20\%$. Note, that $R$ corresponds to the outermost radius, $R_{B}$ or $R_{D}$ depending on the DM distribution for DM core or DM halo formation, respectively. The upper panel shows curves for different values of the coupling constant and two different values of boson mass: $m_{\chi}=400$ MeV (solid curves) and $m_{\chi}=100$ MeV (dashed curves). The lower panel demonstrates an effect of boson mass variation $m_{\chi}=(100-500)$ MeV, while the coupling constant is fixed at $\lambda=\pi$. The green dashed curve shows a DM core-halo transition for which the outermost radius changes from $R_{B}$ to $R_{D}$  (see details in the text).}
    \label{M-R,mass,lambda}
\end{figure}

\begin{figure}[!h]
    \centering
    \hspace{-0.5cm}
    \includegraphics[width=3.2in]{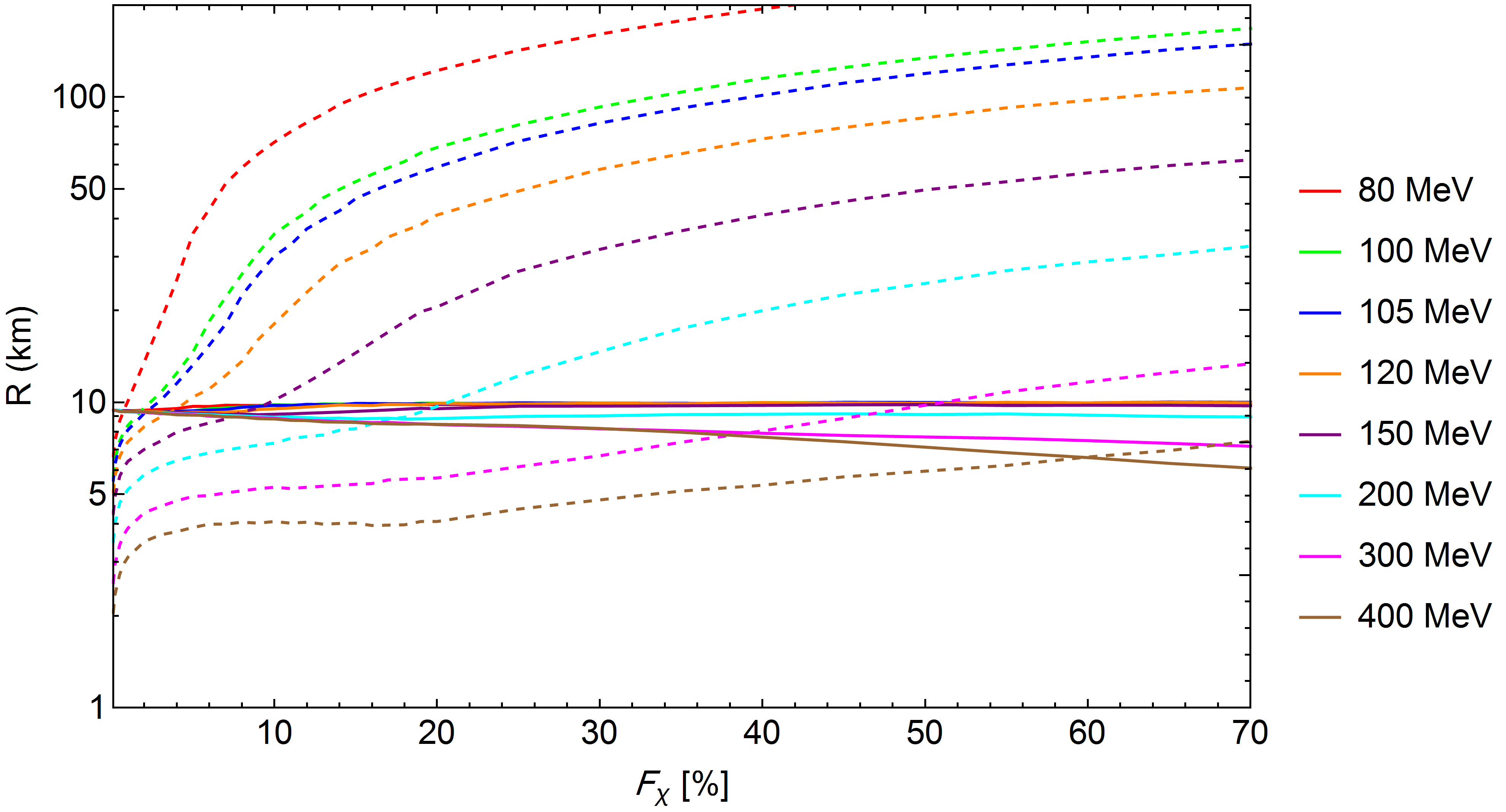} 
    \hspace{.2cm}
    \includegraphics[width=3.4in]{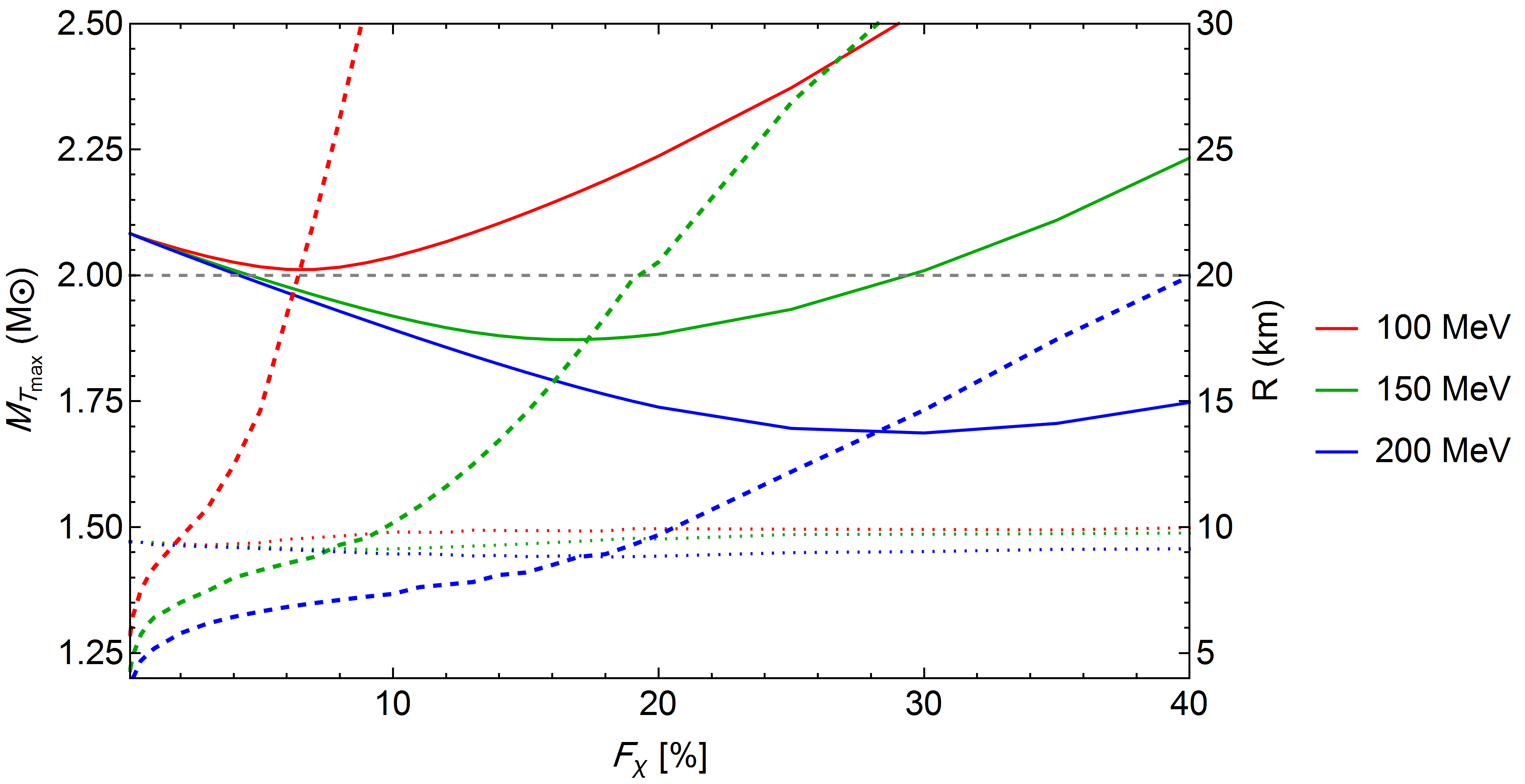}
    \caption{Upper panel: radii of  BM and DM components as a function of DM fraction are presented for different boson mass. $R_B$ and $R_D$ are depicted by solid and dashed curves, respectively. Lower panel: maximum total gravitational mass of DM admixed NSs (solid curves) as a function of DM fraction. On the right side of vertical axis, radii of BM (dotted curves) and DM (dashed curves) components are shown. Gray horizontal dashed line represents $2M_{\odot}$ constraint. Calculations were made for $\lambda=\pi$.}
    \label{FxRM}
\end{figure}

In Fig. \ref{M-R,mass,lambda} we show an effect of different values of $m_{\chi}$ and $\lambda$ on the maximum mass and profile of the M-R relation of DM admixed NSs for a fixed DM fraction $20\%$. In the lower panel, it is shown that a decrease of $m_{\chi}$ leads to an increase of the maximum mass. We find that the star's radius grows very drastically for lower DM masses compared to heavy masses, due to the fact that the outermost radius of the star in the former case is determined by $R_D$ (DM EoS), while the latter one is defined by $R_B$ (BM EoS). For $m_{\chi}=100$ MeV (red dashed curve in the lower panel of Fig. \ref{M-R,mass,lambda}) for which a DM halo forms around a baryonic NS, the total maximum mass increases. On the other hand for $m_{\chi}>200$ MeV a DM core forms inside a NS leading to a decrease of the total mass compared to a pure BM star. As an intermediate regime, we want to point out on the blue dashed curve in the lower panel of Fig. \ref{M-R,mass,lambda} obtained for $m_{\chi}=200$ MeV, $F_{\chi}=20\%$ and $\lambda=\pi$ which shows a reduction of the maximum mass although  a DM halo is formed (see Fig. \ref{FxRM} for more details). 

\begin{figure}[!h]
    \centering
    \includegraphics[width=3.2in]{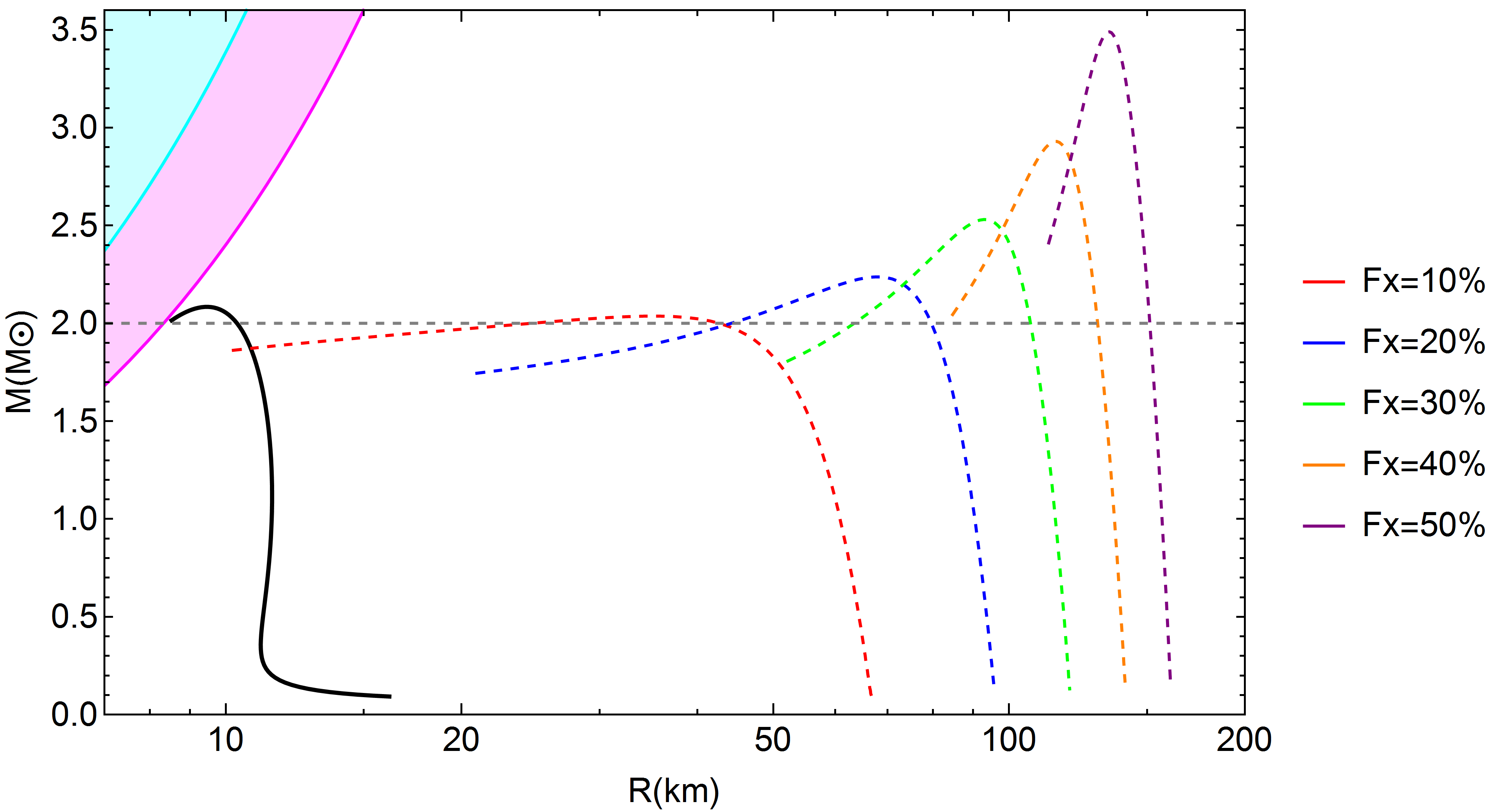}  \includegraphics[width=3.2in]{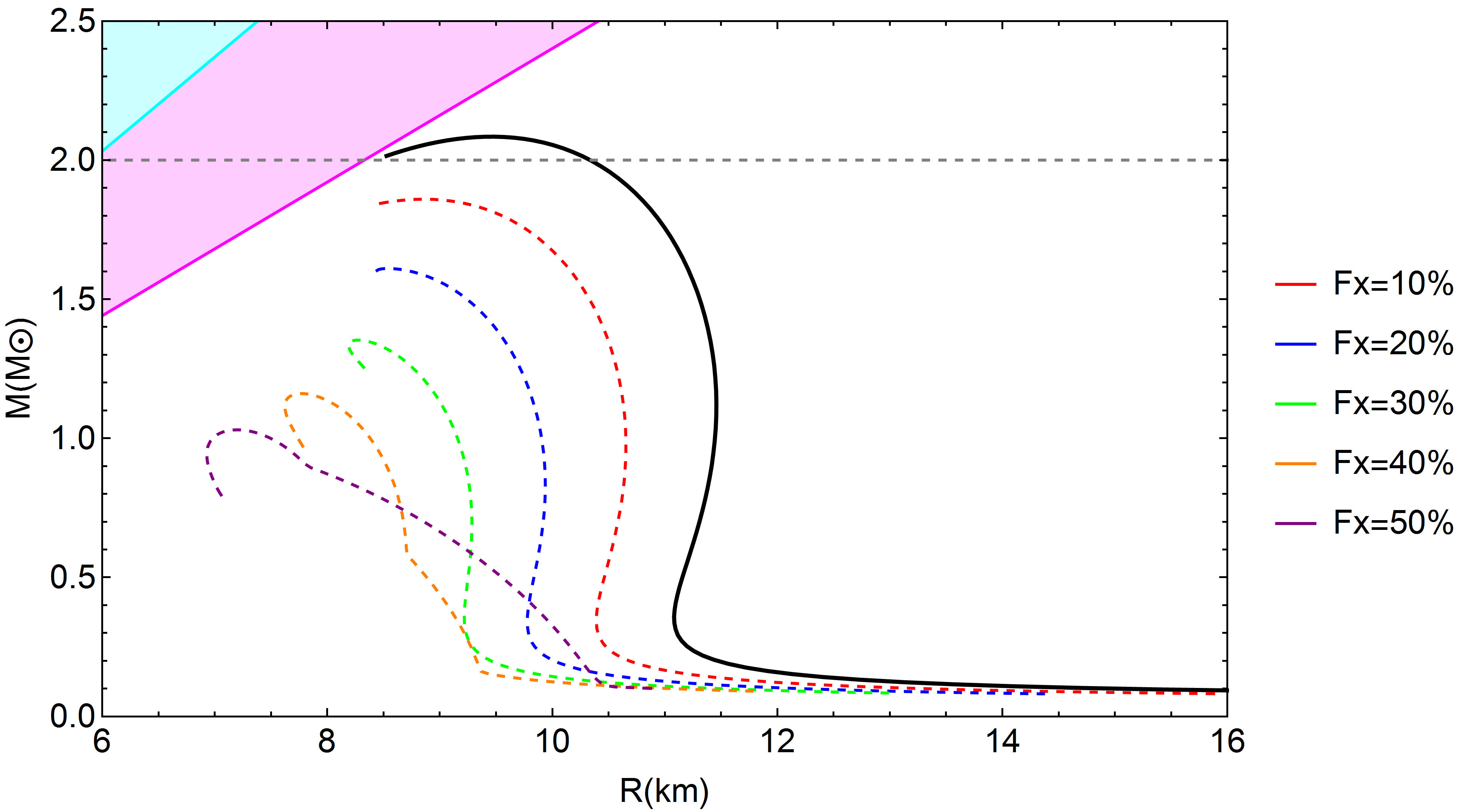}
    \label{fig:event-electron}
    \caption{M-R profiles of DM admixed NSs calculated for various DM fractions and a fixed value of the self-coupling constant $\lambda=\pi$. The upper and lower panels correspond to $m_{\chi}=100$ MeV and $m_{\chi}=400$ MeV, respectively. Here R is the outermost radius defined by either BM or DM component.
    }
    \label{M-R,fraction}
\end{figure}

\begin{figure}[!h]
    \centering
    \includegraphics[width=3.5in]{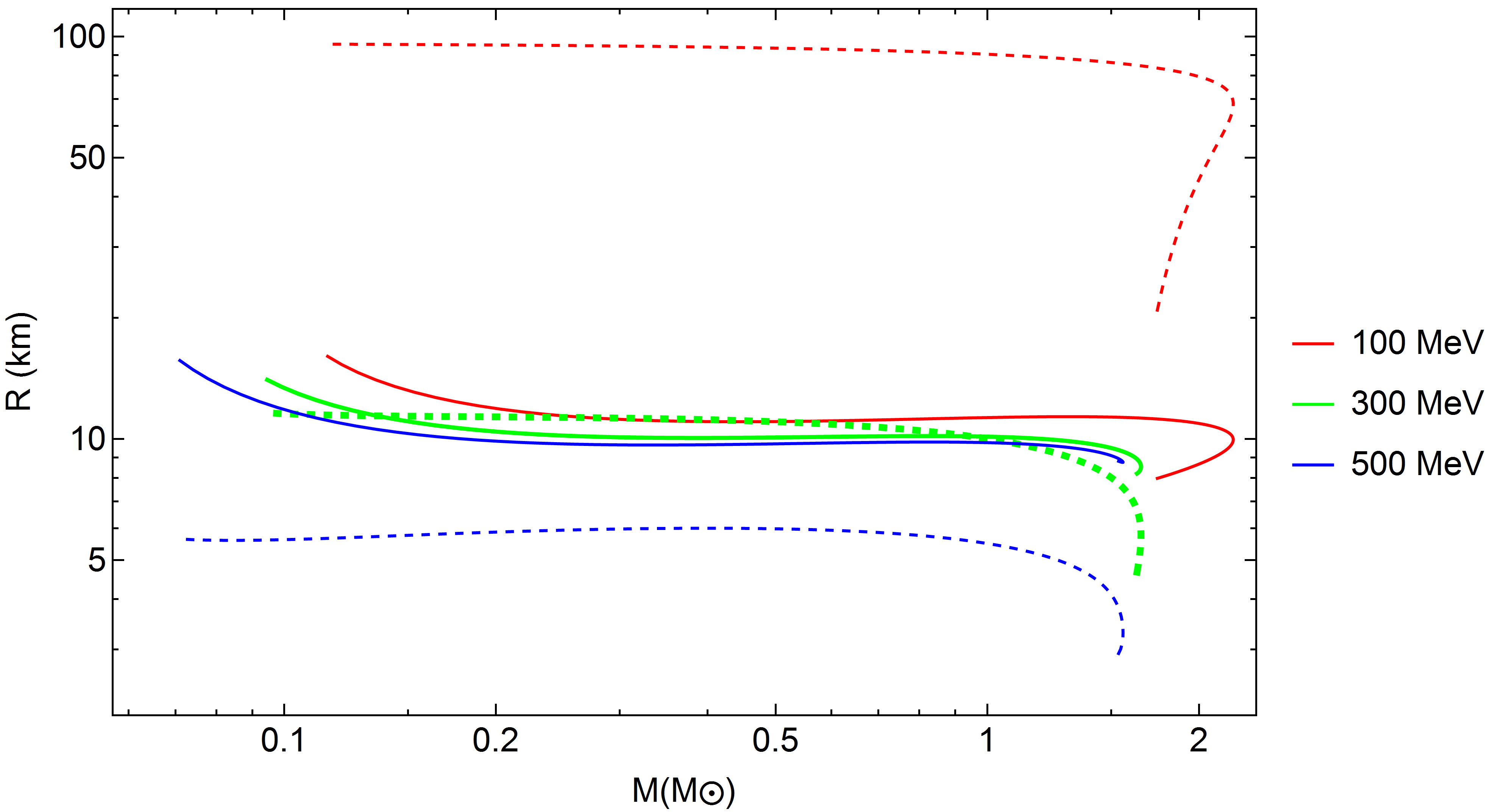}  
    \caption{Radius of DM and BM components separately as a function of 
    the total gravitational mass obtained for $\lambda=\pi$, $F_{\chi}=20\%$ and three different values of boson mass. The radius of BM component ($R_{B}$) is shown by solid curves and DM component ($R_{D}$) is depicted by dashed curves. Red ($m_{\chi}=100$ MeV) and blue ($m_{\chi}=500$ MeV) curves indicate
    two different DM distribution regimes: DM halo and DM core formation, respectively. Green curves represent a DM core - halo transition that corresponds to the M-R profile calculated for $m_{\chi}=300$ MeV in the lower panel of Fig. \ref{M-R,mass,lambda}.}
    \label{M-R phase1}
\end{figure}

\begin{figure}[!h]
    \centering
    \includegraphics[width=3.2in]{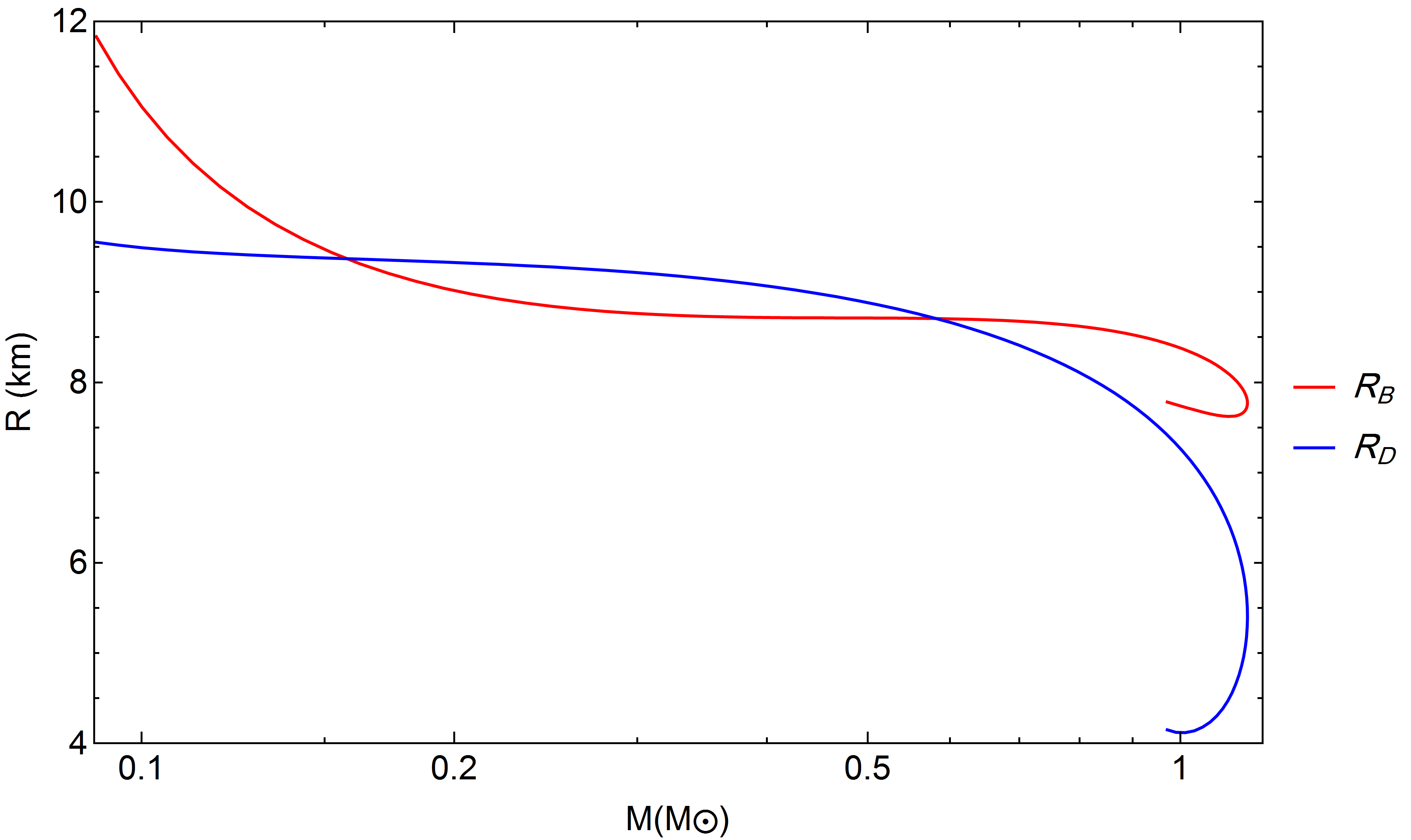}  \includegraphics[width=3.2in]{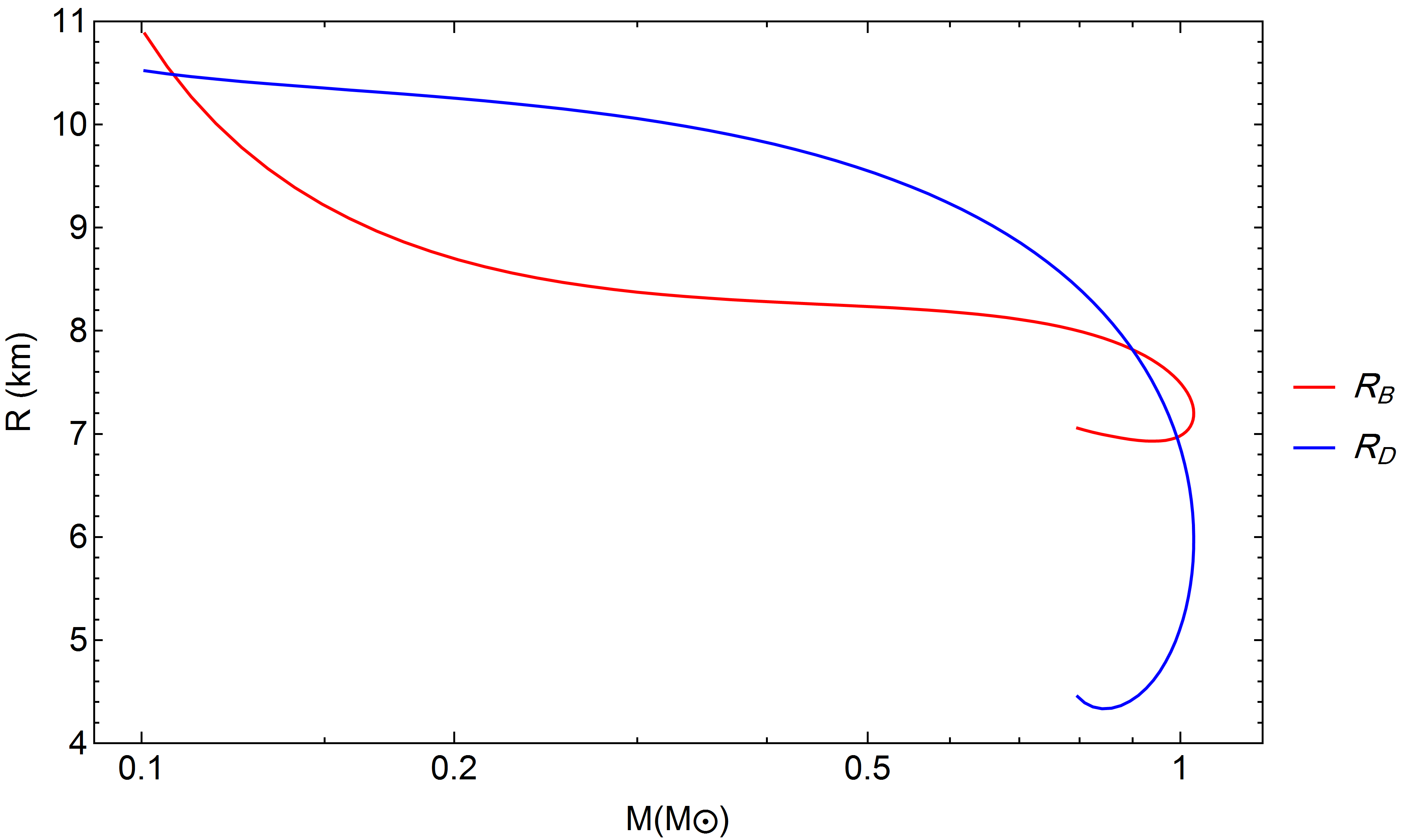}
    \label{fig:event-electron}
    \caption{Radius of  BM $R_{B}$ (red curves) and DM $R_{D}$ (blue curves) components as a function of total gravitational mass are depicted for $F_{\chi}=40\%$ (upper panel), $F_{\chi}=50\%$ (lower panel), fixed value of boson mass $m_{\chi}=400$ MeV  and $\lambda=\pi$. These plots indicate that a transition occurs between two regimes of DM distribution, from a core to a halo structure and vice-versa. Both panels aimed to illustrate and explain a behavior shown on lower panel of Fig. \ref{M-R,fraction}.}   
    \label{M-R phase2}
\end{figure}
In the upper panel of Fig. \ref{M-R,mass,lambda} an effect  of self-coupling constant $\lambda=(0.5,1,1.5,2)\pi$ on the total maximum mass of the compact 
stars is investigated. Here M-R profiles are shown for $m_{\chi}=100$ MeV (dashed curves) as an example of a DM halo and  $m_{\chi}=400$ MeV (solid curves) to illustrate a DM core formation. As you can see on the upper panel of Fig. \ref{M-R,mass,lambda}, DM particles with mass $400$ MeV lead to a decrease of the maximum mass leaving it below $2M_{\odot}$ constraint. At the same time, an increase of $\lambda$ rises  the maximum mass for both DM masses while the star's radius has a small reduction for $m_{\chi}=400$ MeV and  drastically increases for $m_{\chi}=100$ MeV.

We show an impact of DM fraction on the M-R profiles of compact stars in Fig.  \ref{M-R,fraction}. For light DM particles with mass $100$ MeV (see the upper panel) the bigger fraction causes a grow of the maximum mass and radius of stars. For heavier DM particles $m_{\chi}=400$ MeV, we see an opposite behavior that prompt a  reduction of total maximum mass and radius with an increase of fraction (lower panel on Fig.  \ref{M-R,fraction}). 
It is worth mentioning that for all above mentioned cases in which a DM halo is formed ($R_D>R_B$), the visible radius of a star remains to be $R_{B}$.

\begin{figure}[!h]
    \centering
    \includegraphics[width=3.5in]{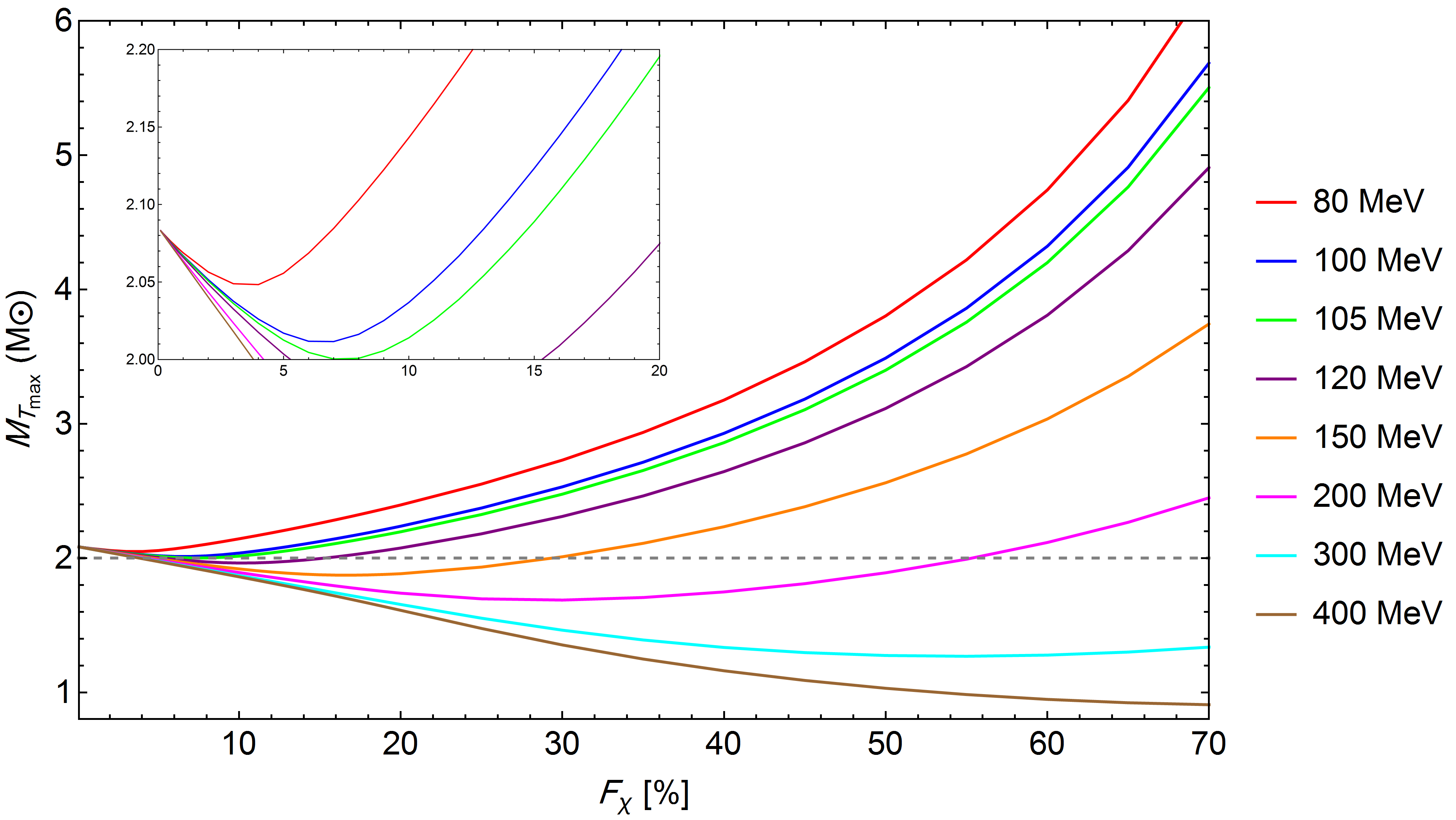}  
    \caption{Maximum total gravitational mass of DM admixed NSs as a function of DM fraction $F_{\chi}$ obtained for a fixed value of coupling constant $\lambda=\pi$ and different values of $m_{\chi}$. It shows that for $m_{\chi}\leqslant105$ MeV and $\lambda=\pi$, the $2M_{\odot}$ constraint is satisfied for all $F_{\chi}$ values.}
    \label{fractionmass}
\end{figure}

\begin{figure}[!h]
    \centering
    \includegraphics[width=3.5in]{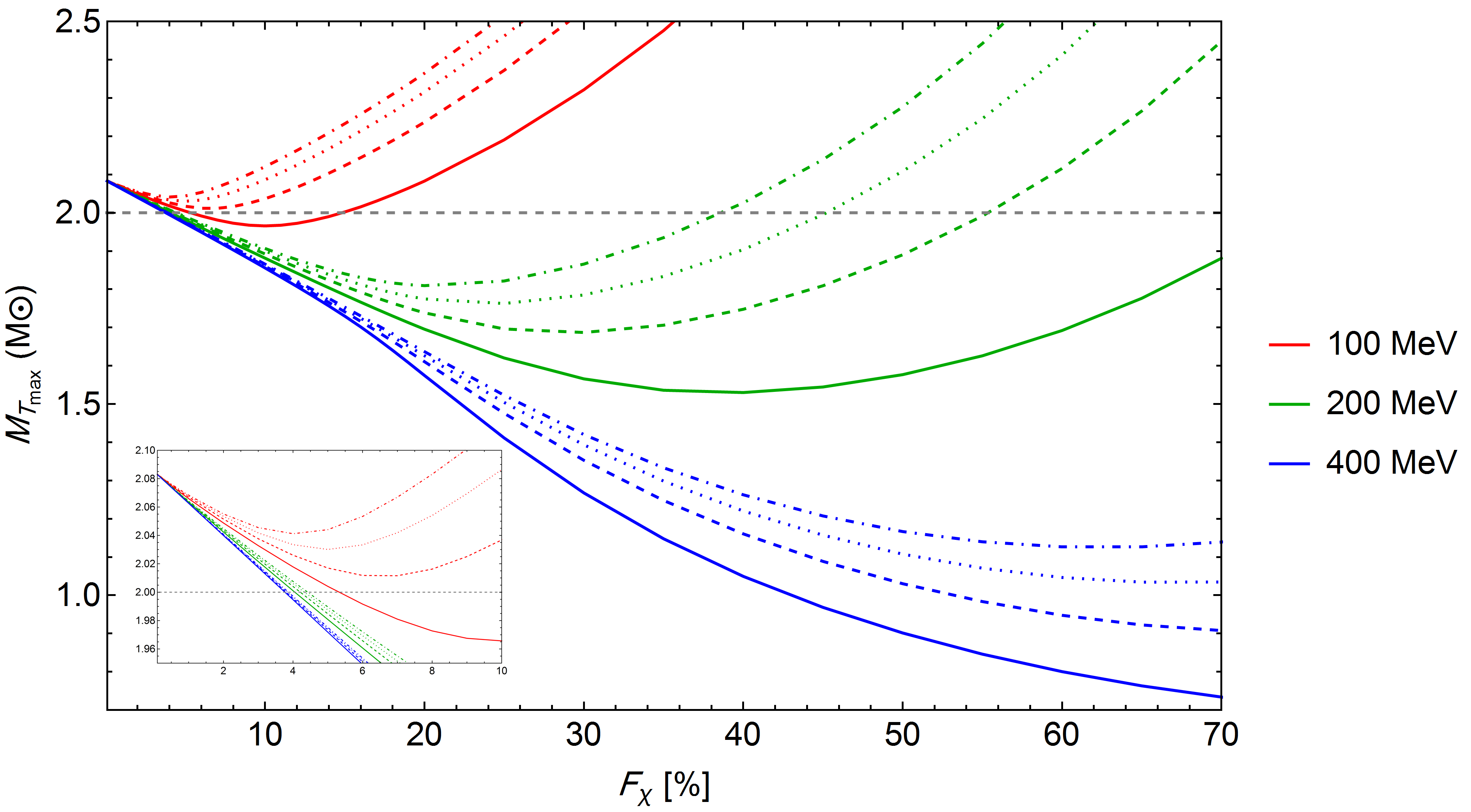}
    \caption{Maximum total gravitational mass of DM admixed NSs as a function of DM fraction $F_{\chi}$ for $m_{\chi}=(100,200,400)$ MeV and $\lambda=0.5 \pi$ (solid curves), $\lambda= \pi$ (dashed curves), $\lambda= 1.5\pi$ (dotted curves),
    $\lambda= 2\pi$ (dot-dashed curves).}
    \label{fractionmass2}
\end{figure}

\begin{figure}[!h]
    \centering
    \includegraphics[width=3.5in]{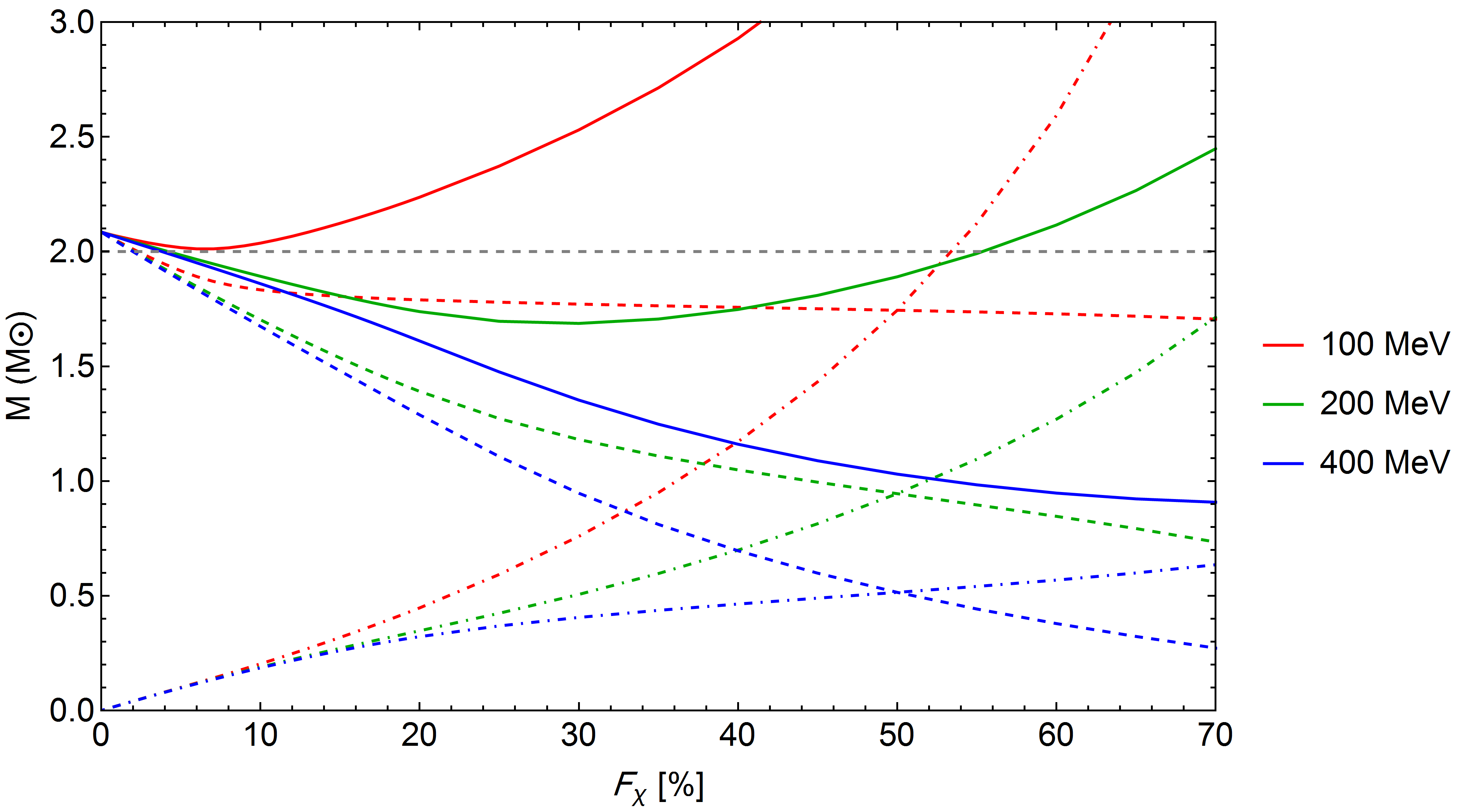}  
    \caption{The maximum total gravitational mass, mass of BM ($M_{B}$) and DM ($M_{D}$) components are separately presented  by solid, dashed and dot-dashed curves, respectively, for $\lambda=\pi$.  The gray dashed line indicates $2M_{\odot}$ limit.}
    \label{fractionmass3}
\end{figure}

 We see here that in agreement with previous studies the existence of a DM core  decreases the maximum stable mass and the corresponding minimum radius while  the formation of a DM halo increases these quantities \cite{Leung:2011zz,Nelson_2019,PhysRevD.102.063028,Ellis:2018bkr}. 
Moreover, we report a new interesting behavior presented on the lower panel of Fig.  \ref{M-R,mass,lambda} for  $m_{\chi}=300$ MeV and $F_{\chi}=20\%$ and on the lower panel of Fig. \ref{M-R,fraction} for $m_{\chi}=400$ MeV and $F_{\chi}=40\%$, 50\% which occurs due to a DM core - halo transition.  In fact, for a given $m_\chi$, $\lambda$ and $F_\chi$ values the outermost radius of the object may interchange between $R_B$ and $R_D$ along the M-R profile for DM admixed NSs. To clarify this new interesting feature, we plotted the radius of BM and DM components separately (see Figs. \ref{M-R phase1}-\ref{M-R phase2}).
Thus, Fig. \ref{M-R phase1} for $m_{\chi}=300$ MeV (green dashed curve) shows that for an intermediate mass range the radius of the DM component exceeds the BM one, $R_D>R_B$, while for the low and high mass tails the baryonic component has a larger radius. Fig. \ref{M-R phase2} illustrates a similar transition for particles with $m_\chi=400$ MeV and high DM fractions $40\%$ and $50\%$. Along the M-R relation $R_D<R_B$ for the low mass stars, $R_D>R_B$ in the intermediate region of total gravitational masses, and  $R_D<R_B$ for massive stars. According to our best knowledge, this is a new feature never reported before, we named it a DM core - halo transition.

To have a clear understanding on how the results depend on DM model parameters, we present a behavior of the maximum total gravitational mass of DM admixed NSs in a wide range of DM fractions in Fig. \ref{fractionmass}. This figure shows that considering $\lambda=\pi$,   (i) for $m_{\chi}\leq 105$ MeV, the maximum mass is always above $2M_{\odot}$ for any DM fraction. The limiting masses for different self-coupling constants are given as ($\lambda, m_{\chi}$)= ($0.5\pi$, 88.4 MeV), ($1.5\pi$, 116 MeV) and ($2\pi$, 125 MeV). (ii) For a DM mass range between  $105$ MeV and $200$ MeV the maximum total mass decreases for low DM fractions, and goes below $2M_{\odot}$. However, by increasing $F_\chi$, the maximum total mass after reaching to a local minimum gradually increases above $2M_{\odot}$ for high DM fractions. This behaviour is a signature of a core to halo transition  induced by variation of amount of DM fraction.  (iii) For bosons with masses of about $m_{\chi}\gtrsim 300$ MeV, we clearly see a DM core formation inside NSs leading to a reduction of the total mass by increasing the DM fraction. For massive DM particles and  high fractions, we  see a small rise of the total maximum mass, however it never reaches to  $2M_{\odot}$ limit even for a pure DM star. 

The effect of different values of self-coupling constant $\lambda=(0.5,1,1.5,2)\pi$ on the total maximum mass as a function of DM fraction is depicted in Fig. \ref{fractionmass2} for  $m_{\chi}$=100, 200 and 400 MeV. This figure shows that the total maximum mass grows by increasing $\lambda$ and  the DM fraction at which $M_{T_{max}}$ crosses $2M_{\odot}$ line has a strong dependence on $\lambda$. Fig. \ref{fractionmass3} illustrates a contribution of BM (dashed curves) and DM (dot-dashed curves) components to  the maximum total gravitational mass (solid curves) of DM admixed NSs. As you can see, in contrast to  $M_{B}$, $M_{D}$ is increasing with $F_{\chi}$, however, the variation rates of these two masses are considerably different for a DM halo (e.g, $m_{\chi}=100$ MeV) and for a DM core (e.g, $m_{\chi}=400$ MeV).
We see from  Fig. \ref{FxRM} that depending on the  DM masses for a fixed $\lambda=\pi$, a DM halo starts to  form at a specific fraction of DM when $R_D>R_B$. The upper panel shows the variation of  radii of the DM component $R_{D}$ (dotted curves) and the BM one $R_{B}$ (solid curves) as a function of DM fraction for different $m_{\chi}$. It indicates that $R_B\approx 10$ km, while $R_{D}$ gradually increases toward larger values. The condition $R_{D}\approx R_{B}$ satisfies when a DM halo appears. On the lower panel of Fig. \ref{FxRM}, the total maximum mass (solid curves), $R_{D}$ (dashed curves) and $R_B$ (dotted curves) are presented in a single plot. It can be seen that a DM halo is appeared for $m_{\chi}=100$ MeV, 150 MeV at $F_{\chi}<10\%$ and for $m_{\chi}=200$ MeV at $F_{\chi}<20\%$. However, the total maximum mass of DM admixed NS starts to increase at higher DM fractions compared to the one of halo formation.

\section{An Effect of Bosonic DM on Tidal Deformability}
\label{sectides}

In this section, we analyse an effect of  bosonic SIDM  on the tidal deformability $\Lambda$ of a DM admixed NS. The 
quadrupole tidal distortion $Q_{ij}$ in terms of external tidal tensor $\mathcal{E}_{ij}$   can be parametrized as follows  
\begin{eqnarray}\label{2tidal}
Q_{ij}=\frac23 k_2 R^5 \mathcal{E}_{ij}=\lambda_{t} \mathcal{E}_{ij}\, ,
\end{eqnarray}
where  $k_2$ is the tidal Love number which can be calculated from the TOV equations \cite{Hinderer:2007mb,Postnikov:2010yn}. Therefore, the tidal deformability $\lambda_t$ strongly depends on the star's EoS.

Unlike $\lambda_{t}$ which has dimension, dimensionless tidal deformability $\Lambda$ can be defined as
\begin{eqnarray}\label{tidall}
\Lambda=\frac{\lambda_{t}}{M^5}=\frac23 k_2 \left(\frac{R}{M}\right)^5\,.
\end{eqnarray}
Here R and M are the radius and mass of a compact star, $k_2$ is calculated by the method presented in Refs. \cite{Hinderer:2007mb,Hinderer:2009ca,Postnikov:2010yn} as 
\begin{eqnarray}\label{k2}
k_2&=&\frac{8C^5}{5}(1-2C)^2[2+2C(y-1)-y] \\ \nonumber
&&\times \{2C[6-3y+3C(5y-8)] \\ \nonumber
&&+4C^3[13-11y+C(3y-2)\\ \nonumber
&&+2C^2(1+y)]+3(1-2C)^2[2-y \\ \nonumber
&&+2C(y-1)]\ln(1-2C)\}^{-1}\,,
\end{eqnarray}
here $C=M/R$ is the compactness and  $y$ is related to the quadrupolar perturbed metric function. It is determined at the star's surface  $y\equiv y(r) |_{r=R}$ through solving the following differential equation with the appropria\-te boundary conditions \cite{Hinderer:2007mb}
\begin{eqnarray}\label{1tidal}
ry'(r)+y(r)^2+y(r)e^{\lambda(r)}\left\{1+4\pi r^2 \right. \\ \nonumber
\left.\left[p(r)-\epsilon(r)\right]\right\}+r^2 Q(r)&=&0.
\end{eqnarray}
Assuming a spherically symmetric star, the metric functions $\lambda(r)$ and $\nu(r)$ are given by
\begin{eqnarray}
e^{\lambda(r)}&=&\left[1-\frac{2M(r)}{r}\right]^{-1},\\
\frac{d\nu}{dr}&=&\frac{2}{r}\left[\frac{M(r)+4\pi p(r) r^3}{r-2M(r)}\right].
\end{eqnarray}
By using EoSs for BM and DM as inputs and setting the initial condition $y(0)=2$ \cite{PhysRevC.95.015801,Postnikov:2010yn}, the values of $y$, $k_2$ and $\Lambda$ can be calculated by simultaneously solving the TOV equations and Eq. (\ref{1tidal}).
For a two fluid system composed of DM and BM, the parameters $\epsilon$, $p$ and $M$ are defined as
\begin{eqnarray}
p=\sum_{i} p_{i},\ \  \epsilon=\sum_{i} \epsilon_{i}, \ \ M=\sum_{i} M_{i}, \ \ \text{i=BM, DM}.~~~~~
\end{eqnarray}
The parameter $Q(r)$ (see Appendix B of Ref. \cite{Das:2020ecp}) is given  by 
\begin{eqnarray}
Q(r)=4\pi e^{\lambda(r)}\left[5\epsilon(r)+9p(r)+\sum_i 
\frac{ \epsilon_i(r)+ p_i(r)}{d p_i/d\epsilon_i} \right]  \nonumber \\
-6\frac{e^{\lambda(r)}}{r^2}-(\nu'(r))^2.\ ~~
\end{eqnarray}
Note that in a  DM admixed NS, $y$ and $C$ and, therefore, $k_2$ should be determined at the outermost radius of the object. In other words, the tidal deformability parameter is sensitive to the gravitational radius which might be different from the visible radius of the star. For a DM halo $R=R_{D}$ and for a DM core  $R=R_{B}$. Meanwhile, stiffness and softness of the EoS affects the tidal deformability through $k_2$ parameter.

\begin{figure}[!h]
    \centering
    \hspace{-0.2cm}
    \vspace{0.3cm}
    \includegraphics[width=3.3in]{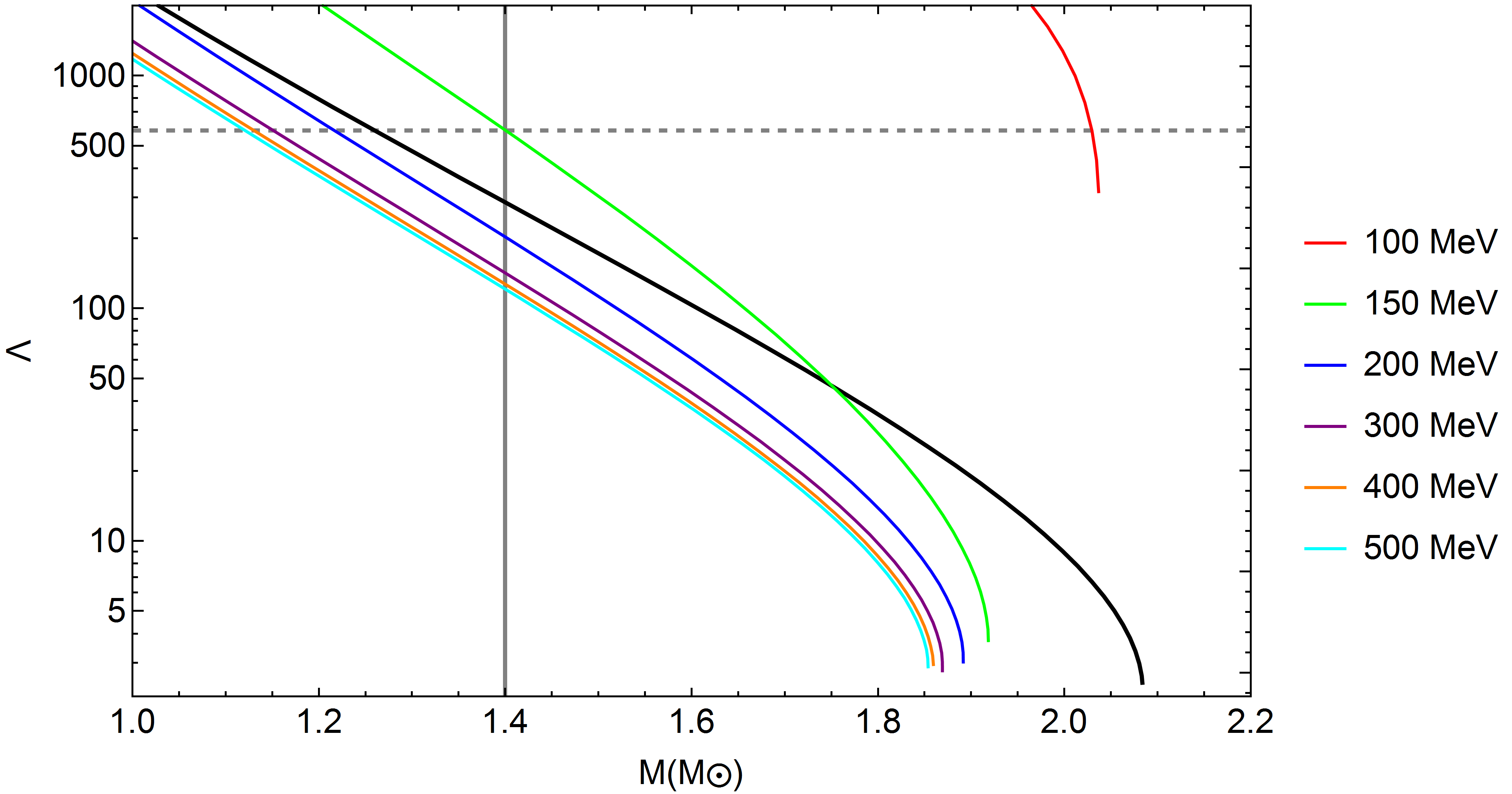} 
    \includegraphics[width=3.3in]{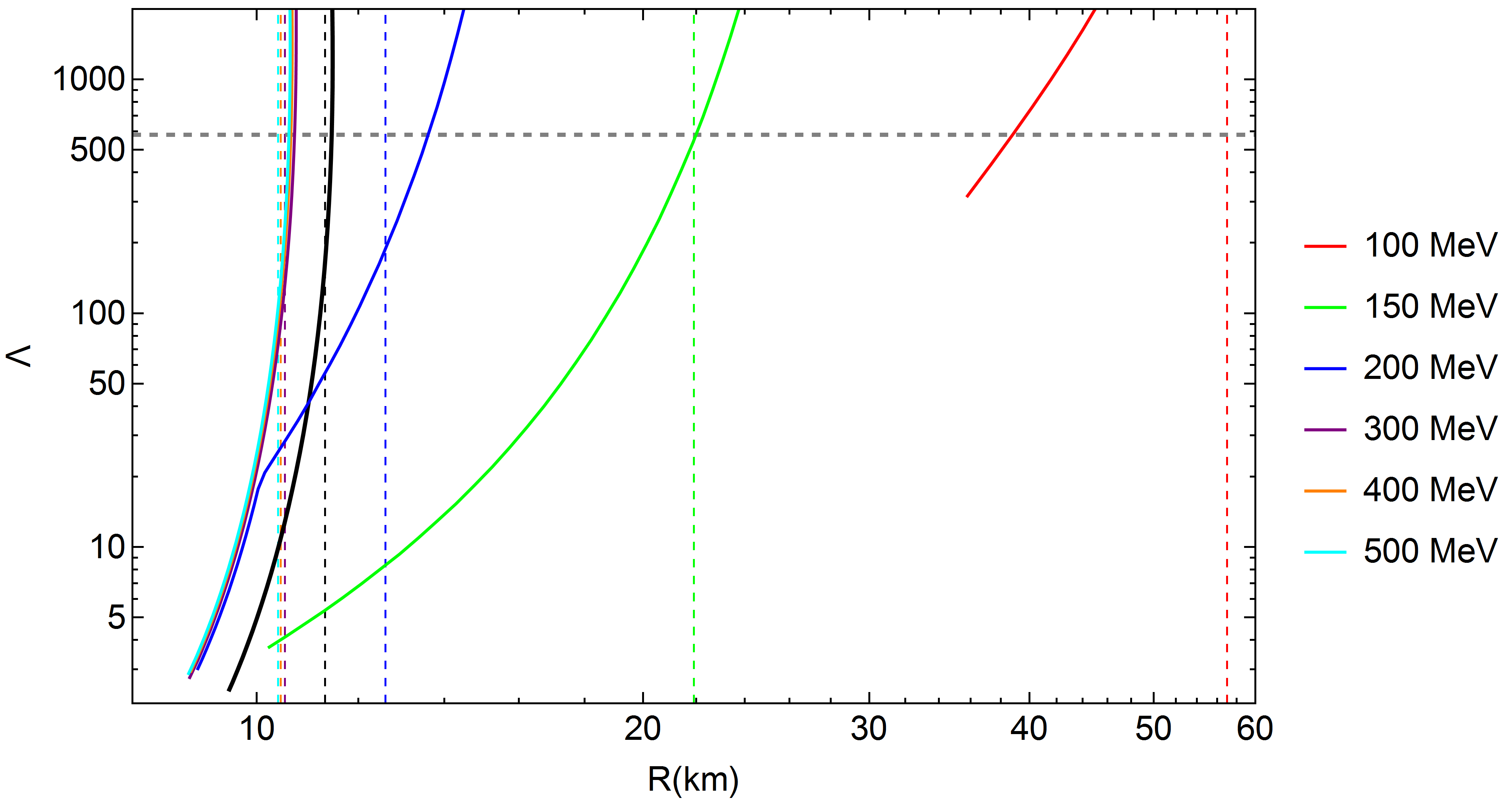}
    \caption{Dimensionless tidal deformability ($\Lambda$) as a function of total gravitational mass (upper panel) and outermost radius (lower panel) presented for various boson masses. Calculations are performed for fixed  $\lambda=\pi$ and $F_{\chi}=10\%$. The black solid curve corresponds to pure BM stars (without DM), gray solid and dashed lines denote $M=1.4M_{\odot}$ and $\Lambda=580$, respectively. On the lower panel each vertical line corresponds to $R_{1.4}$ obtained for different values of boson mass.}
    \label{Lam-M1}
\end{figure}

\begin{figure}[!h]
  \centering
  \hspace{-0.2cm}
    \vspace{0.3cm}
    \includegraphics[width=3.3in]{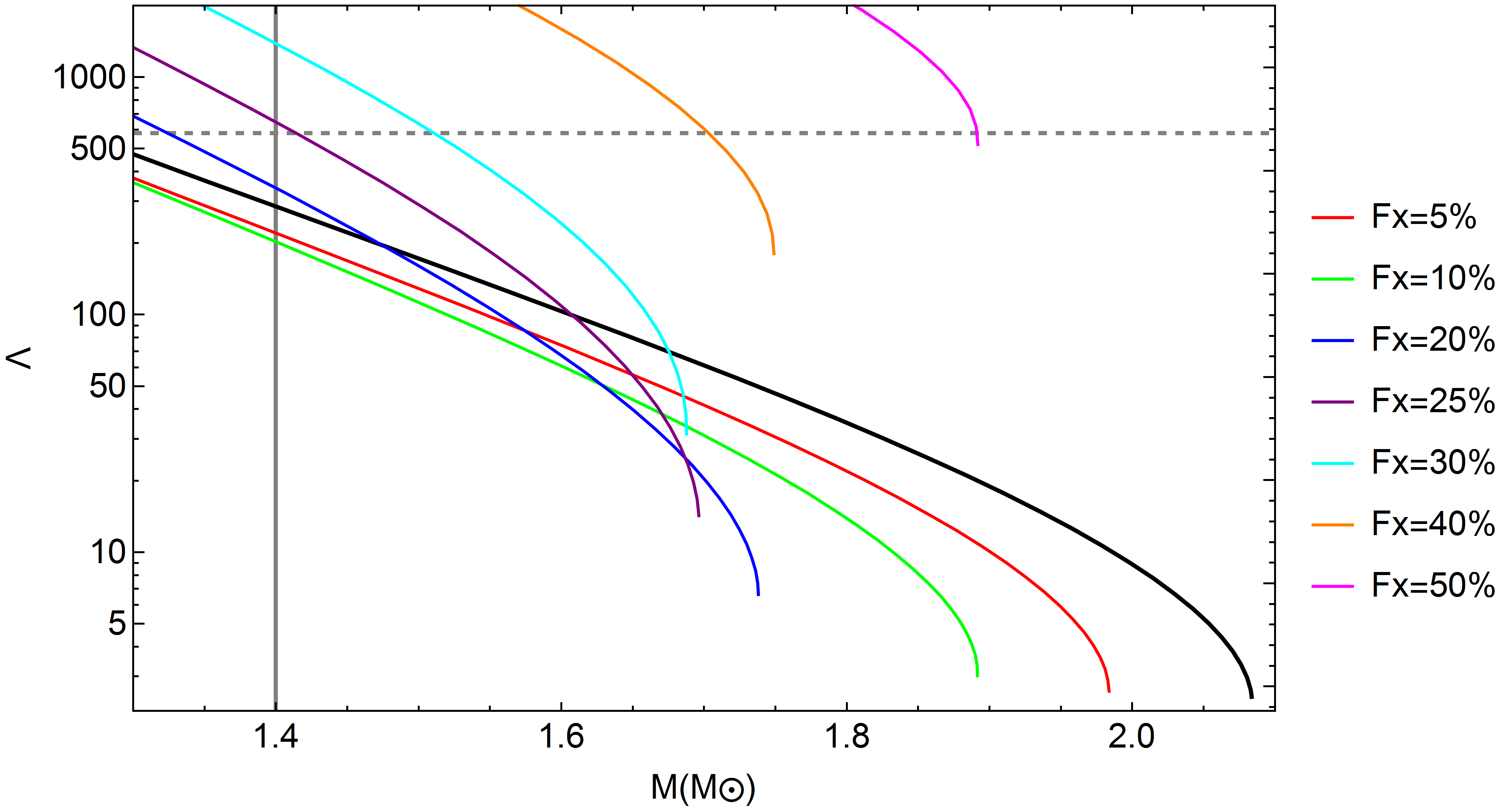}  
    \includegraphics[width=3.3in]{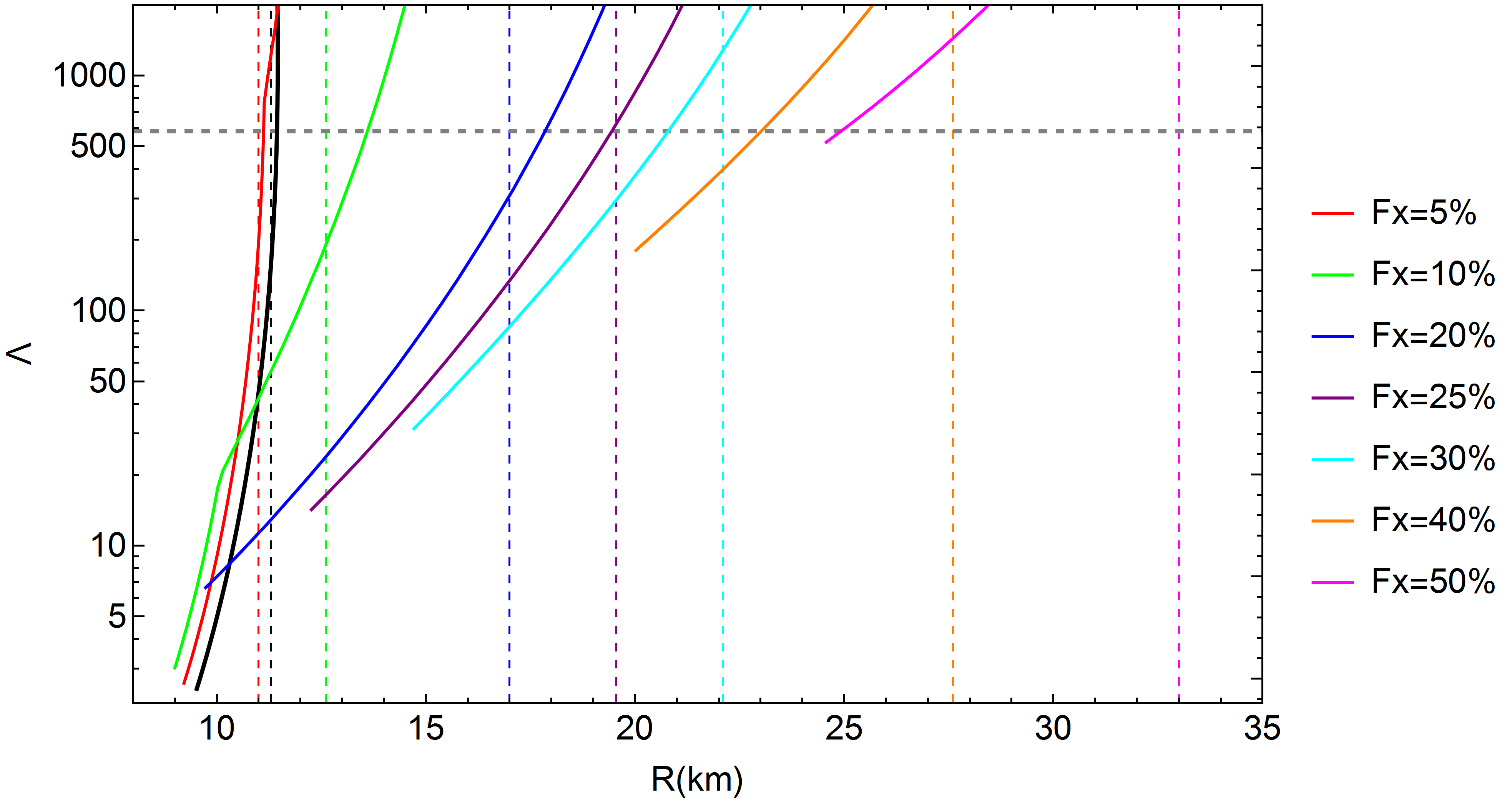}
    \caption{ The same as Fig. \ref{Lam-M1}, except for various DM fractions at  fixed $\lambda=\pi$ and $m_{\chi}=200$ MeV.  On the lower panel each vertical line corresponds to  $R_{1.4}$  obtained for different values of $F_{\chi}$.}
    \label{Lam-M2}
\end{figure}

\begin{figure}[!h]
    \centering
    \hspace{-0.2cm}
    \vspace{0.3cm}
    \includegraphics[width=3.3in]{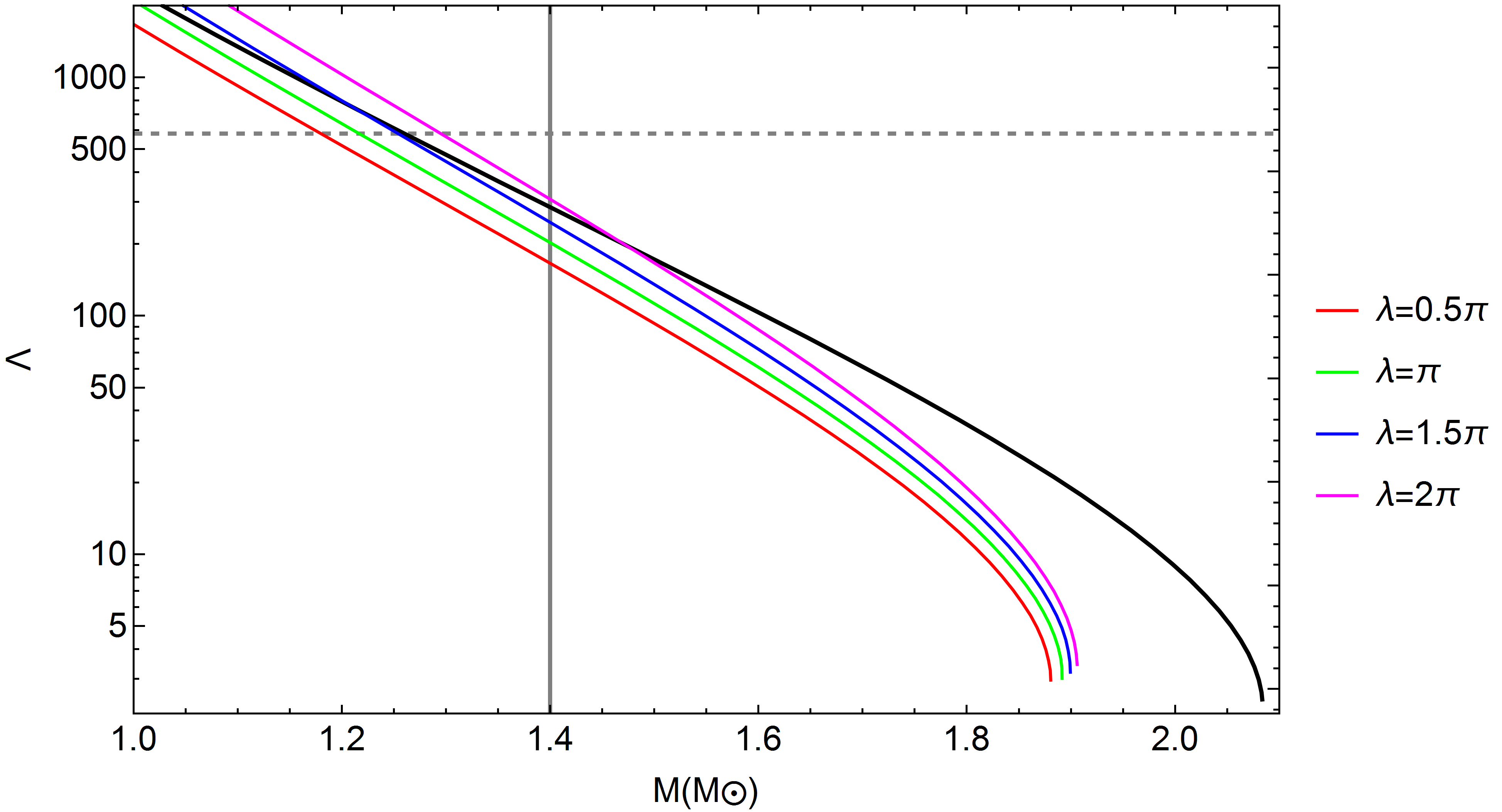}  \includegraphics[width=3.3in]{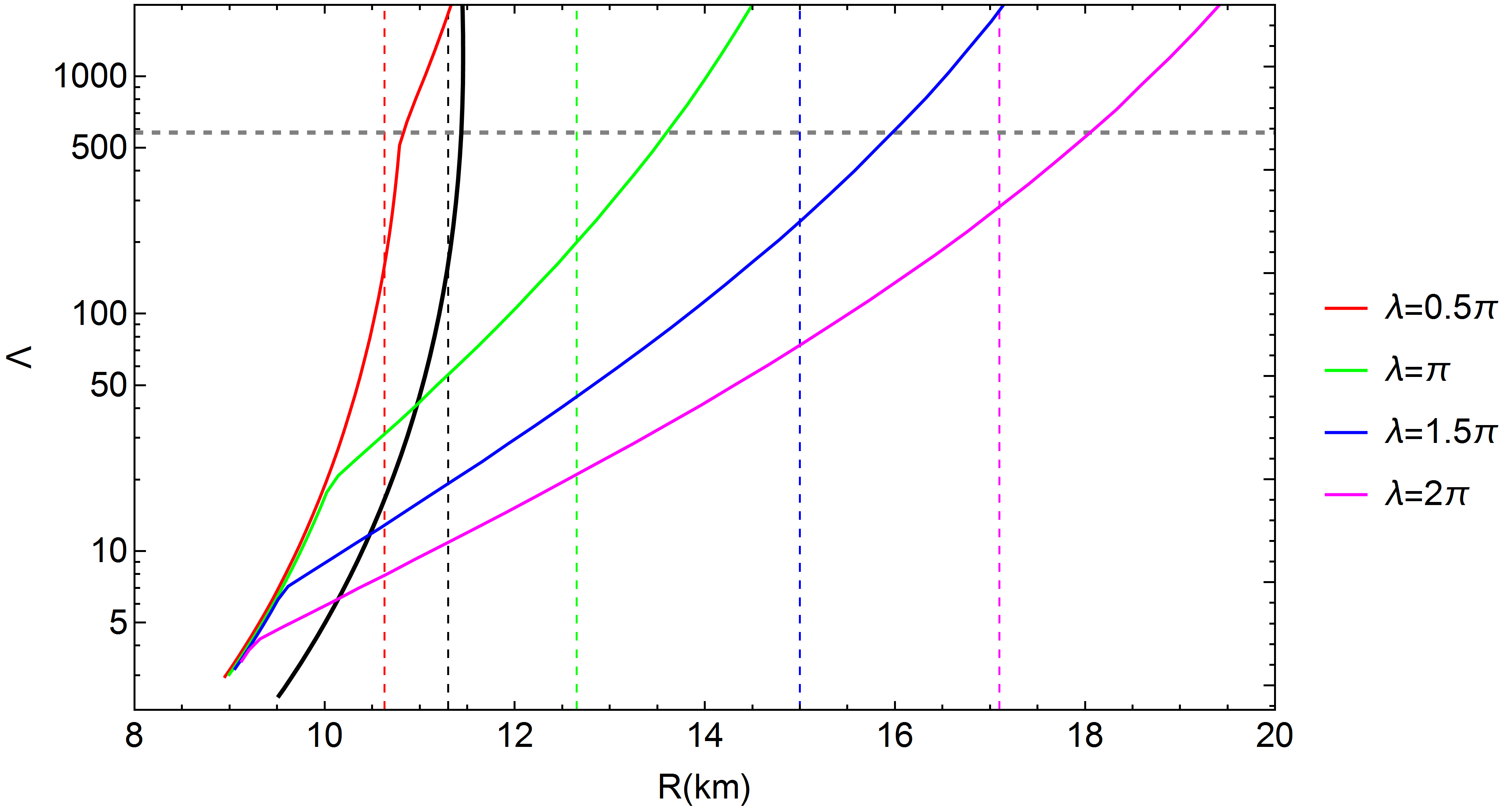}
    \caption{The same as Figs. \ref{Lam-M1}-\ref{Lam-M2}, but for various self-coupling constants at fixed $F_{\chi}=10\%$ and $m_{\chi}=200$ MeV values. On the lower panel each vertical line corresponds to the $R_{1.4}$ obtained for different values of $\lambda$.}
    \label{Lam-R}
\end{figure}

In the following, we investigate  the effect of the bosonic SIDM  distributed either as a DM core or a DM halo on the dimensionless tidal deformability parameter. The dependence of $\Lambda$ on the total gravitational mass and radius of DM admixed NSs is shown in Figs. \ref{Lam-M1} -\ref{Lam-R} for different values of $m_\chi$, $\lambda$ and $F_{\chi}$. In these figures the gray horizontal dashed  lines indicate the LIGO/Virgo upper bound $\Lambda_{1.4}=580$  \cite{Abbott:2018exr}, the gray solid vertical lines show $M_{T}=1.4M_{\odot}$ and the colored dashed vertical lines stand for $R_{1.4}$ radius for the corresponding model parameters.

The tidal deformability calculated for the pure baryonic IST EoS (see Sec. \ref{secist}) is denoted by the solid black curve in Figs. \ref{Lam-M1} -\ref{Lam-R}. As you can see, its $\Lambda_{1.4}$ value is well below the LIGO/Virgo constraint.  Thus, within the IST EoS we are able to model both core and halo distributions without violating the tidal deformability constraint. It is related to the fact that the presence of a dense DM core or an extended halo effectively leads to decrease or  increase of $\Lambda$, respectively.  However for those BM EoSs for which $\Lambda_{1.4}>580$, in order to be compatible with GW170817 tidal limit, the presence of DM mainly as a core component is allowed.

In fact, a general profile of the tidal deformability in Figs. \ref{Lam-M1} -\ref{Lam-R}  is a decreasing (increasing) function of the total gravitational mass (radius). From  Eq. (\ref{tidall}), we can see that $\Lambda$ is a function of $R/M$, and, therefore, its lowest value is associated with the maximum mass and/or minimum radius of the DM admixed NS. The main reason that the tidal deformability increases when a DM halo forms around a NS and reduces when a DM core forms inside it, is related to a strong dependence of $\Lambda$ on the stellar radius (the outermost radius) and mass through Eq. (\ref{tidall}). 

The effect of varying the boson mass $m_\chi$ on the tidal deformability is illustrated in  Fig. \ref{Lam-M1} for  $\lambda=\pi$ and $F_{\chi}=10\%$. As you can see, a decrease of the DM particle's mass leads to an increase of $\Lambda$.  Such a behavior is in agreement with our understanding, since for light DM particles a DM halo tends to form. It consequently causes a growth of the outermost radius of the object giving rise to higher values of the tidal deformability parameter. On the lower panel of Fig. \ref{Lam-M1}, it is shown how $R_{1.4}$  grows as $m_\chi$ becomes  smaller. At the same time in the upper panel, the tidal deformability for $m_{\chi}< 200$ MeV lies above the curve for the IST EoS (black curve). However, when a DM core is formed inside a NS (e.g, $m_{\chi}\geq300$ MeV), $\Lambda$ value drops below the one for the IST EoS. Note that a change in the behavior of the tidal deformability curve as a function of R for $m_{\chi}=200$ MeV  is an indication of a DM core - halo transition which has been extensively discussed in Sec \ref{secmass}. During a transition the outermost radius of the star switches between $R_B$ and $R_D$.

Fig. \ref{Lam-M2} shows how DM fraction affects the tidal deformability of DM admixed NSs at the fixed value of boson mass $m_\chi=200$ MeV and $\lambda=\pi$. Higher DM fractions correspond to a DM halo formation leading to higher $\Lambda$ values and $\Lambda_{1.4}\leq580$ constraint is fulfilled for $F_{\chi}\leq25\%$. However, low DM fractions give rise to a DM core formation, and, consequently, cause a reduction of the tidal deformability to be below the IST curve. A DM core -  halo transition is among the features that appears by changing $F_\chi$. In  Fig. \ref{Lam-R},  the tidal deformability is calculated for different values of the self-coupling constant between $0.5\pi$ and $2\pi$ for $m_{\chi}=200$ MeV and $F_{\chi}=10\%$. It turns out that higher values of $\lambda$ generate larger $\Lambda$. Note, that a core-halo transition can be observed for all curves in Fig.  \ref{Lam-R}.

It is important to note that in the case of a DM halo formation for relatively light DM particles with $m_{\chi}\leq100$ MeV, radius of DM component $R_D$ (see Figs. \ref{M-R,mass,lambda} - \ref{M-R,fraction}) can reach even above 100 km, which leads to significant enhancement of the value of tidal deformability. 

At the moment, an analysis of the inspiral phase of NS-NS coalescence does not include hydrodynamic simulations, and therefore it is limited to a case of finite separation between the stars. To stay in agreement with the present GW analysis, we restrict ourselves to  $R_{D}\le75$ km to prevent an overlap of DM halos which corresponds to lower frequencies detectable by Ad. LIGO \cite{Nelson_2019,Ellis:2018bkr}.

\begin{figure}[!h]
  \centering
  \hspace{-0.2cm}
    \includegraphics[width=3.6in]{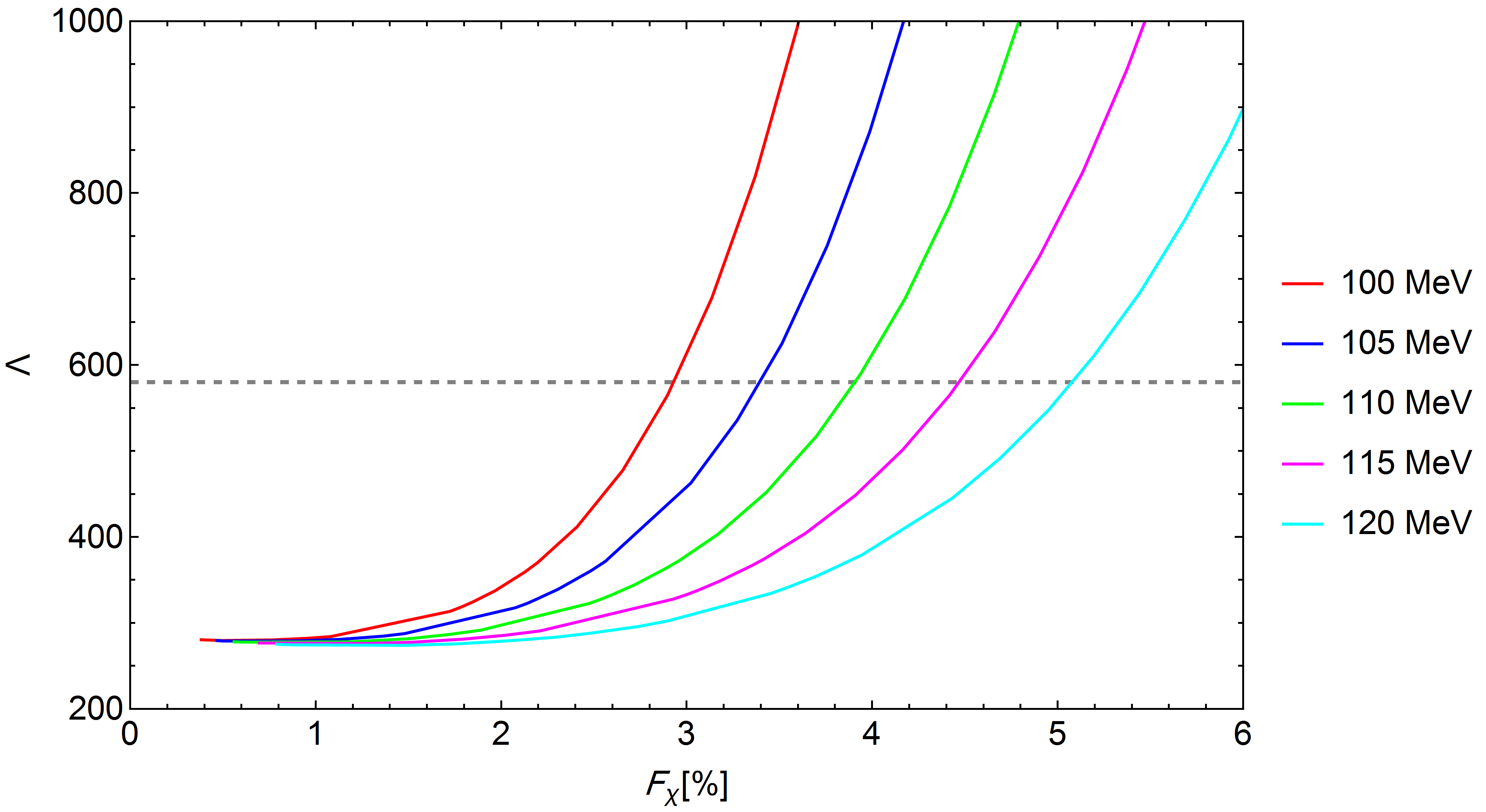}
    \\
    \centering
    \hspace{-0.6cm}
    \includegraphics[width=3.62in]{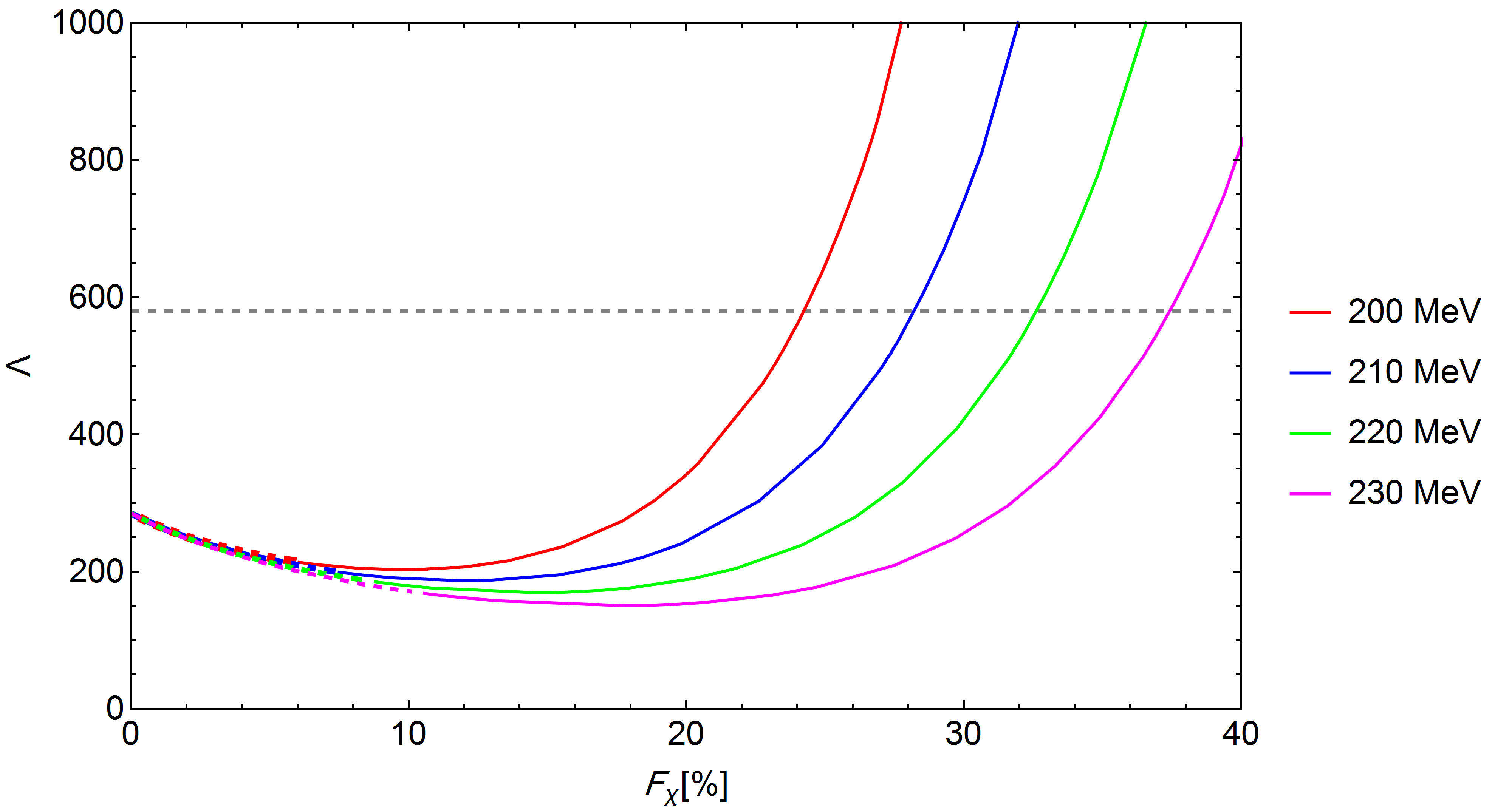} 
    \\
     \hspace{0.7cm}
     \centering
      \includegraphics[width=3.55in]{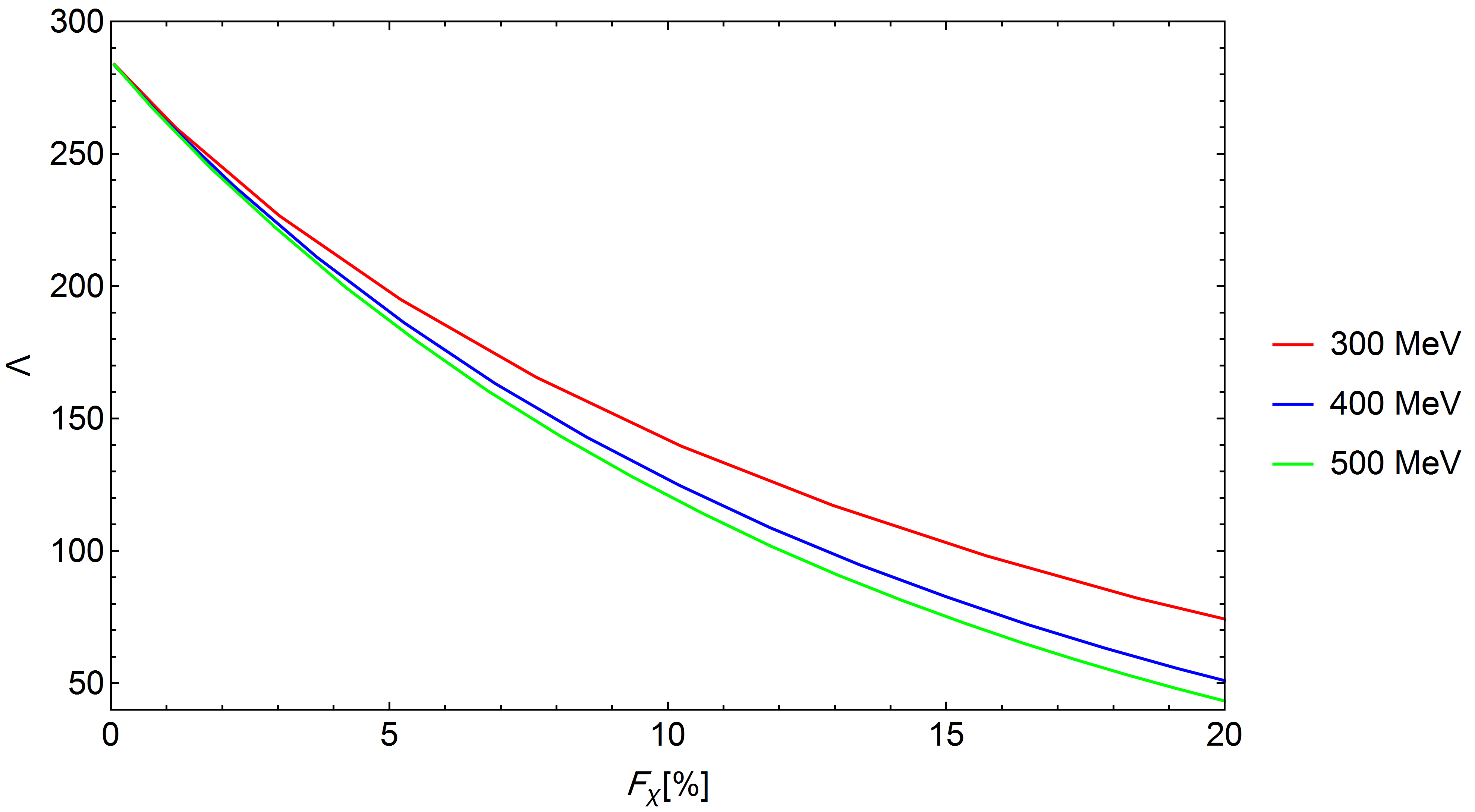}
    \caption{Dimensionless tidal deformability ($\Lambda$) vs. DM fraction plotted for light $m_{\chi}$ at which a DM halo is formed (upper panel),  intermediate $m_{\chi}$ values that show both a DM core (dashed curves) and a DM halo (solid curves) formation (see the middle panel), high $m_{\chi}$ values cause a DM core formation (lower panel). For all panels $\lambda=\pi$ and $M_{T}=1.4M_{\odot}$. Gray dashed line on the first two panels denotes $\Lambda_{1.4}=580$ constraint.}  
    \label{Fx-Lam1}
\end{figure}
\begin{figure}[!h]
    \centering
    \vspace{0.6cm}
    \includegraphics[width=3.35in]{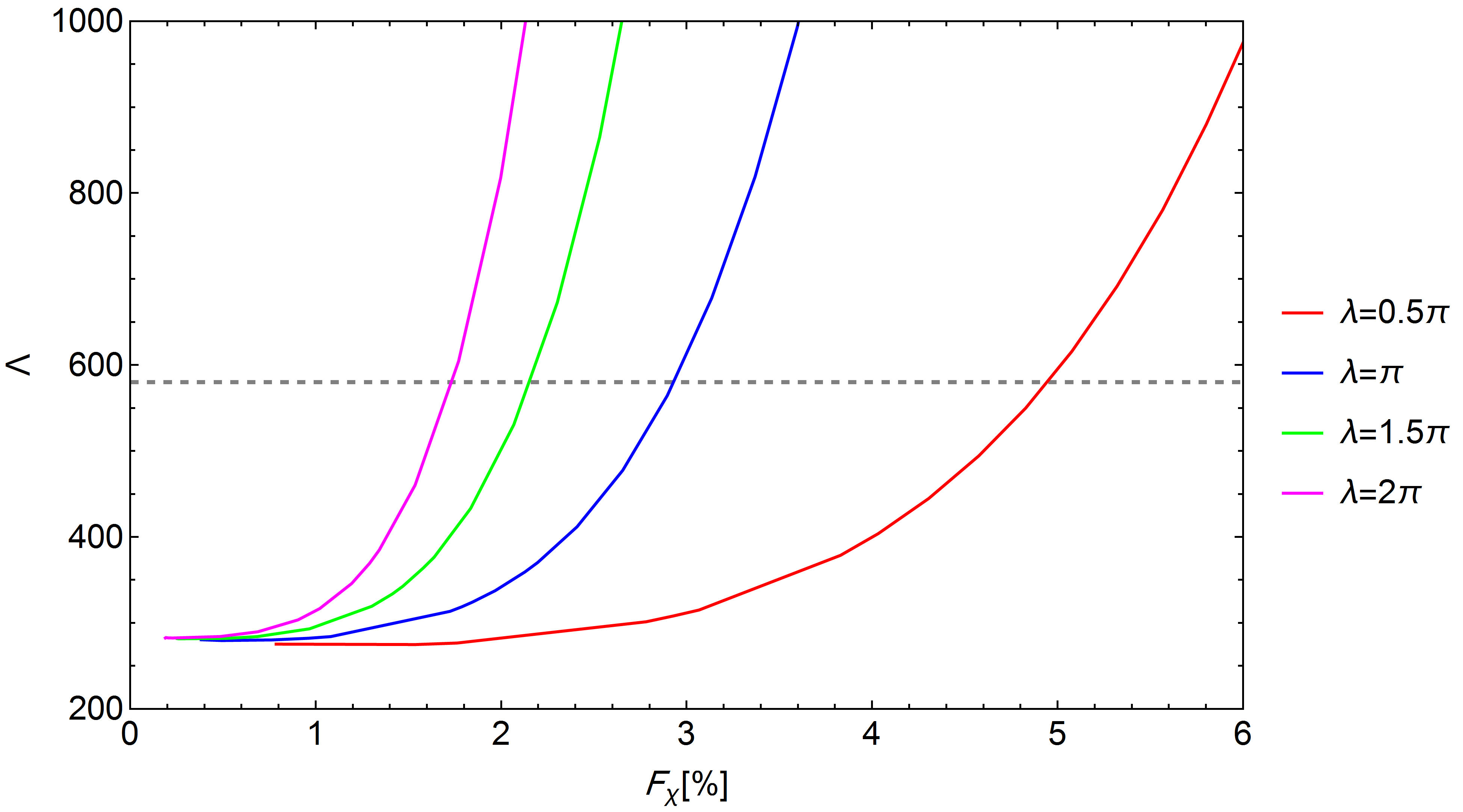}
    \includegraphics[width=3.35in]{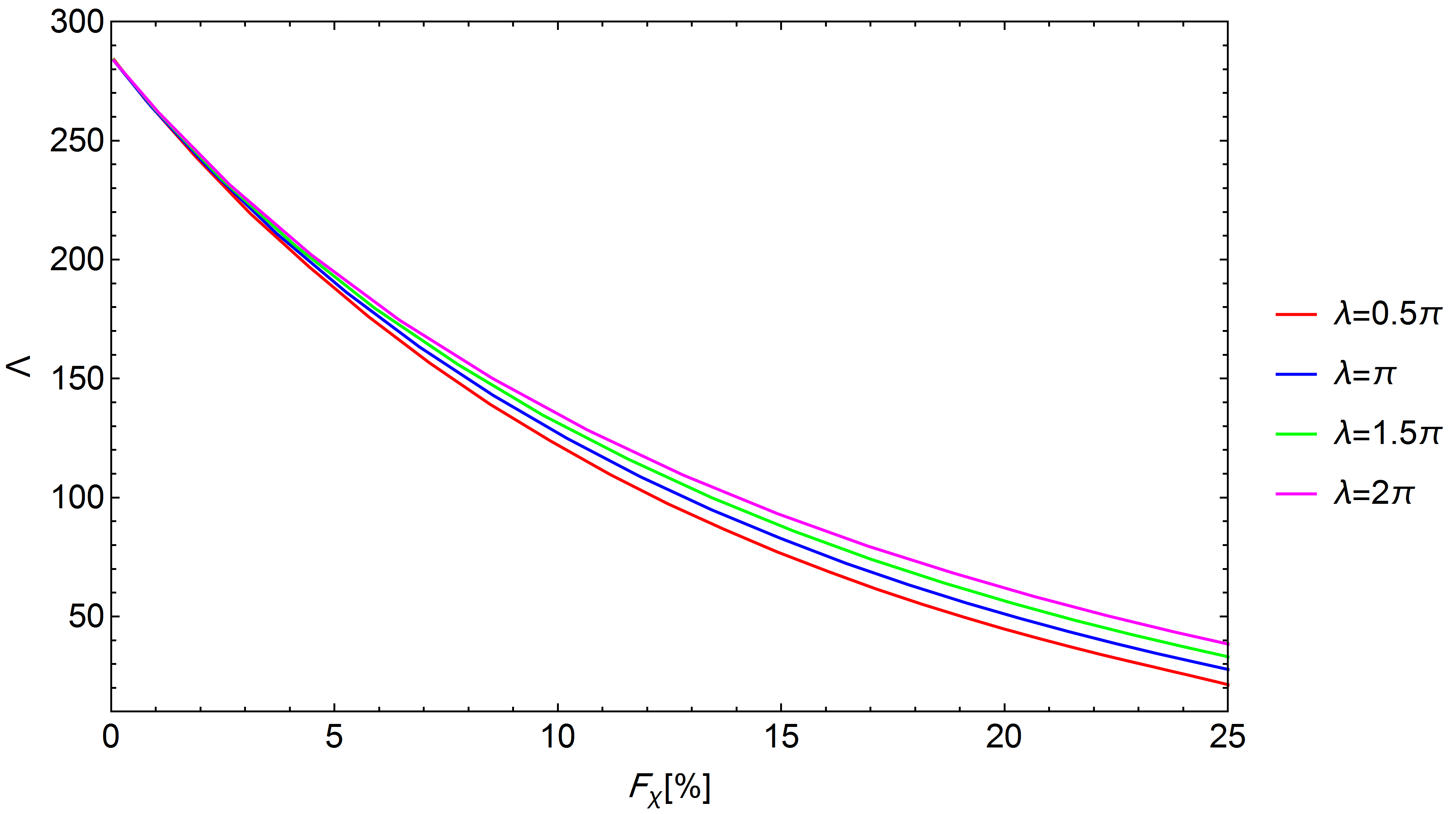}
    \caption{The same as Fig. \ref{Fx-Lam1}, but  for  different values of self-coupling  constant  at fixed $m_{\chi}=100$ MeV (upper panel) and $m_{\chi}=400$ MeV (lower panel). 
    The gray dashed line on the upper panel denotes $\Lambda=580$ constraint for $1.4M_{\odot}$ star.}
    \label{Fx-Lam2}
\end{figure}

To give more insight, taking into account the observable constraint from GW170817  \cite{Abbott_2017,Abbott:2018exr}, in Fig. \ref{Fx-Lam1}, we demonstrate the tidal deformability as a function of $F_{\chi}$ for a fixed value of the total star's mass $1.4M_{\odot}$. On the upper panel, the tidal deformability is an increasing  function of DM fraction for the light DM particles where a DM halo is formed for  $F_{\chi}\leq6\%$. The horizontal gray line indicates $F_\chi$ at which a DM admixed NS satisfies the $\Lambda_{1.4}\leq580$ constraint.
On the middle panel, we consider an intermediate mass range between 200 MeV - 230 MeV for which the tidal deformability behavior shows the features of both a DM core (part of the curve depicted as dashed) and a DM  halo (part of the curve depicted as solid). We see that for  low DM fractions $F_{\chi}\leq 10\%$, DM tends to condensate in the core of a NS and   $\Lambda$  is a decreasing function of $F_{\chi}$. The tidal deformability  reaches minimum at a specific value of DM fraction at which a core to halo transition occurs and afterwards $\Lambda$ starts to grow with $F_{\chi}$.

 On the lower panel of Fig. \ref{Fx-Lam1}, we show three curves calculated for  $m_{\chi}=(300,400,500$) MeV that tend to create a dense DM core inside a NS. In this case the tidal deformability is a monotonically decreasing function, which heavier DM particles cause bigger reduction in $\Lambda$ at a fixed fraction.  Regarding the tidal deformability of NSs admixed with heavy DM particles, we should be careful about the lower tidal limit reported by LIGO/Virgo Collaborations to be $\Lambda_{1.4}=70$ \cite{Abbott:2018exr}. Meanwhile, according to the recent results of NICER \cite{2019ApJ...887L..24M, Raaijmakers_2020,Miller:2021qha} and astrophysical observations \cite{2016ARA&A..54..401O,Raaijmakers:2021uju}, we know that  $R_{1.4}$ is about 12 km, this is an additional reason for cutting the DM core curves at a specific DM fraction.

Finally, for the sake of completeness, the effect of DM self-interaction strength $\lambda$ for $m_{\chi}=100$  MeV (upper panel) and $m_{\chi}=400$ MeV (lower panel) for  $1.4M_{\odot}$ star is presented in Fig. \ref{Fx-Lam2}. 
It turns out that for higher values of the coupling constant an allowed range of DM fractions, consistent with $\Lambda_{1.4}\leq580$ constraint, is decreased (see the upper panel of Fig. \ref{Fx-Lam2}). The lower panel shows that higher $\lambda$ leads to higher tidal deformability  for a fixed DM fraction. In fact the DM EoS for the higher coupling constants is stiffer which causes larger values of the tidal deformability.  Thus, regardless of the DM being distributed in a core or a halo, higher values of $\lambda$ leads to larger $\Lambda$ at fixed $F_{\chi}$ for DM admixed NSs.

In summary, we show that in agreement with the previous studies the DM halo causes an increase of the tidal deformability \cite{Nelson_2019}, while the DM core reduces the tidal deformability compared to the case of pure BM \cite{Ellis:2018bkr}. Interestingly for the first time we show that there is a continues transition between a DM core and a DM halo regimes for different DM masses and fractions. In fact depending on the DM fraction, particles of different masses in sub-GeV range can form either a DM core or a DM halo, as was shown previously in Fig. \ref{FxRM}. The light bosons $m_{\chi}\lesssim 200$  MeV form a DM core  at very small fractions and DM halo at intermediate and large fractions while  heavy bosons $m_{\chi}\gtrsim 300$ MeV lead to a DM halo  at very large values of $F_\chi$ and a DM core at intermediate and small $F_\chi$. 

\section{Constraint on the Fraction and Mass of Dark Matter}
\label{secfrac}

\begin{figure}[h]
    \centering
    \includegraphics[width=3.3in]{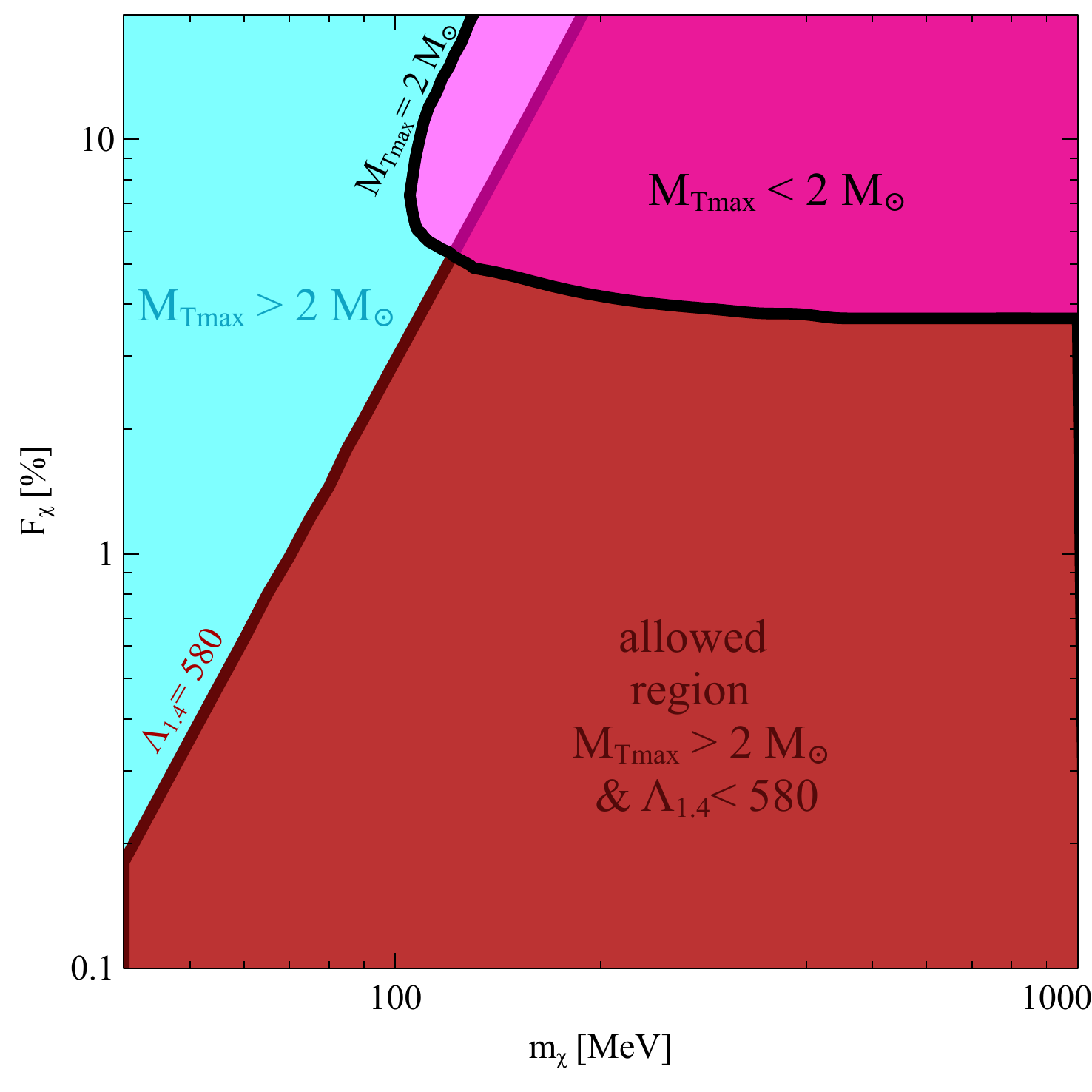}
    \caption{The fraction of DM as a function of its particle mass for $\lambda=\pi$. The black curve represents the maximum total gravitational mass to be equal to $2M_{{\odot}}$. The cyan region is in agreement with $2M_{{\odot}}$ constraint, while the magenta area corresponds to not allowed region of parameters. The dark red line indicates $\Lambda_{1.4}=580$ constraint on tidal deformability. The region below the black curve and on the right from the dark red line is in a full agreement with the heaviest known NSs and LIGO/Virgo constraints.}
    \label{fraction}
\end{figure}

As it was shown in the Secs. \ref{secdistribuition}-\ref{sectides}, the properties of DM admixed NSs depend on three model parameters: boson mass $m_{\chi}$, DM fraction  $F_{\chi}$ and the value of the coupling constant $\lambda$. To make the final conclusion about the allowed range of parameters in a 2D plot, we fixed $\lambda=\pi$ as the medium and most representative value of the coupling constant in our consideration. 
In Fig. \ref{fraction}, the remaining two parameters are plotted with an indication of the total maximum mass and the tidal deformability values. The solid black curve in Fig. \ref{fraction} indicates the values of $F_{\chi}$ and $m_{\chi}$ for which the total maximum gravitational mass equals to  $M_{T_{max}}=2M_{{\odot}}$. Thus, any point below the black curve (cyan region in  Fig. \ref{fraction}) gives $M_{T_{max}}>2M_{{\odot}}$ which is in agreement with the two heaviest observed pulsars \cite{Antoniadis:2013pzd,NANOGrav:2019jur}.
The dark red curve depicts the tidal deformability constraint $\Lambda_{1.4}=580$  \cite{Abbott:2018exr}. Hence, all the parameter space on the right hand side of the red line (dark red region in Fig. \ref{fraction}) yields $\Lambda_{1.4}\leq580$ values.
Note that in a log scale the $\Lambda_{1.4}=580$ constraint is a straight line.
A DM core formation constraints an upper limit of the allowed range of parameters due to  decrease of the maximum mass of DM admixed NS for $m_{\chi}>200$ MeV and $F_{\chi}\lesssim10\%$. However, lighter bosons that form a DM halo around a NS impose a lower limit on the allowed range of parameters, since a DM halo increases the tidal deformability and could exceed the limit reported by the LIGO/Virgo Collaboration.

From  Fig. \ref{fraction}, we see that allowed DM fractions inside NS drastically narrowed down by inclusion the tidal deformability constraint, more specifically for $m_{\chi}\lesssim 70$ MeV it imposes a limit on the amount of DM to be less than $1\%$ of the total NS mass. We can conclude that for sub-GeV bosonic DM, the exi\-sting observational data support low DM fractions below $5\%$. According to our study, the recent measurements for the maximum mass and tidal deformability of NSs are in agreement with the DM admixed NS scenario. The fraction of DM in their interior is compatible with the amount accreted during star's lifetime as well as its possible augmentation that will be discussed in Sec. \ref{accretion}. By applying two observable quantities we set a stringent constraint on the amount of DM inside NSs. Further narrowing down the DM fractions is possible by considering simultaneous measurements of mass and radius, e.g, by ongoing NICER observations \cite{Riley:2021pdl, Miller:2021qha, Raaijmakers:2021uju}.

It is worth mentioning that Fig. \ref{fraction} shows the results for one value of the coupling constant $\lambda=\pi$. For the higher values of $\lambda$ the $M_{T_{max}}=2M_\odot$ curve will be lifted up, increasing the range of fractions compatible with the heaviest observed NSs. 
On the other hand, the $\Lambda_{1.4}=580$ curve will be shifted to the right, limiting the allowed values of $F_{\chi}$ consistent with  the LIGO/Virgo constraint. A detailed analysis of the above mentioned effects and a scan over different values of the coupling constants will be subject of a following paper.

\section{Dark Matter Accumulation Regimes}
\label{accretion}

An important question we want to address at this  section is related to how compact stars can contain and accumulate DM in their interior. The capturing rate of DM by NSs 
depends on the local density of DM, the DM-BM  scattering cross section, and the DM mass \cite{Bramante:2013hn,Bell:2020obw,Bell:2020jou}. If the DM decay or annihilation is permitted, the number of DM particles in a NS could be depleted. The most plausible scenario for the presence of DM in NSs is its accumulation throughout different stages of star's lifetime. In this regard, four main evolution phases should be considered: (a) progenitor, (b) main sequence star, (c) supernova explosion with formation of a proto-NS, and (d) equilibrated NS.
Depending on the distance of a star from the Galactic center the local DM density varies significantly, and consequently the amount of accreted DM \cite{Navarro:1995iw,1965TrAlm...5...87E,Ruffini_2015,Arg_elles_2018, 2020Univ....6..222D, PhysRevD.102.063028,Ciancarella:2020msu}. The total accreted mass in a spherically symmetric accretion scenario for a typical NS with $M=1.4M_{\odot}$ and  $R=10$ km is given by
\begin{equation}
M_{acc}\approx 10^{-14} \left( \frac{\rho_{dm}}{0.3 \ GeV/cm^{3}}\right) \left(\frac{\sigma_{\chi n}}{10^{-45} cm^{2}} \right)\left( \frac{t}{Gyr}\right) M_{\odot},
\end{equation}
where $\rho_{dm}$ is the local density of DM,  $\sigma_{\chi n}$ is the nucleon-DM elastic cross section and t denotes the age of the NS \cite{Kouvaris:2010vv,Kouvaris:2013awa}. While the DM density in the Solar system is about $0.3$~GeV$/$cm$^{3}$, it can reach at most  $\sim  (10^{11}-10^{12})~$GeV$/$cm$^{3}$ near the center of the Galaxy for certain DM profiles \cite{Bertone:2005hw,Merritt:2003qk,Freese:2008hb}. It can be shown that the accreted mass can be varied from $\sim 10^{-13}-10^{-14} M_{\odot}$ to $\sim 10^{-5}-10^{-8} M_{\odot}$ \cite{PhysRevD.102.063028,DelPopolo:2019nng,Baryakhtar:2017dbj,Bramante:2015dfa,Guver:2012ba}.
Moreover, DM production in the NS interior might be an additional effective mechanism which should be taken into account. For instance, due to high baryon density in the core of compact stars, during a supernova explosion or a binary NSs merger a creation of DM particles from nucleons could be triggered, whereas a major part of DM could be created and trapped inside a NS \cite{Nelson_2019,Ellis:2018bkr}.
High DM fractions \cite{Li_2012,Leung:2011zz,Leung:2012vea,Ciancarella:2020msu,Ciarcelluti:2010ji,Sandin_2009}  cannot be easily obtained during a typical star's lifetime from normal accretion processes considering only a smooth spatial distribution of DM in the Galaxy. However, high DM factions inside compact stars can be acquired by accounting for additional scenarios: (i) clumps of DM were present at the early stages of the Universe 
forming seeds/accretion centers for BM, this process may lead to a DM admixed NS even with dominant contribution of DM. In fact, instead of accretion of DM onto ordinary NS, one may assume accretion of ordinary matter onto a pre-existing dark
core \cite{Ellis:2018bkr,Goldman_2013,Ciarcelluti:2010ji,Deliyergiyev_2019}. (ii) Since the density of DM at a certain distance from the Galactic center  in a first approximation is homogeneous, a NS can pass through a region in space with locally high DM density leading to an accretion of a large amount of DM \cite{Sandin_2009,DelPopolo:2020pzh,Deliyergiyev_2019,Li:2012ii,2020Univ....6..222D,Profumo:2006bv,Berezinsky:2013fxa,Berezinsky:2014wya}.  (iii) One might speculate a formation of a stable compact object composed of ADM as a dark star \cite{Kouvaris_2015,Maselli:2017vfi,Eby:2015hsq}. Hence, NS  could  capture DM from the dark star companion \cite{Ellis:2018bkr,Goldman_2013,PhysRevC.89.025803} or   a merger like event may occur \cite{Ciarcelluti:2010ji,Gresham:2018rqo,Sandin_2009}. Regarding all above three possible scenarios, we showed that the existence of high DM fractions is tightly constrain by joint observations of the tidal deformability and the maximum mass.

\section{Conclusion and Remarks}
\label{secconclusion}
In this paper, we have investigated the possible effects of   bosonic ADM with repulsive self-interaction on the compact star properties including its radius and  gravitational mass. We have shown that depending on  DM model parameters such as boson mass $m_\chi$, self-coupling constant $\lambda$ and DM fraction $F_\chi$,  DM can be distributed either in a core or a halo. The impact of various DM distribution regimes on observable quantities e.g, the maximum total gravitational mass  and the tidal deformability has been considered. We found that DM condensed  in the core of a NS leads to decrease of the total gravitational mass, radius and the tidal deformability compared to a typical baryonic NS. On the other hand, the presence of  DM particles in the halo around the NS increases those observa\-ble quantities. A rich phenomenology of the scenario presented in this article  allows a transition between the DM core and halo for different particle's masses and fractions. During the DM core - halo transition, the outermost radius of the object interchanges from the radius of BM to DM component. This leads to some new features in mass-radius profile and the tidal deformability-radius behavior of the DM admixed NS. 

As our main result, we show a combined analysis of the observational data for the heaviest observed NSs and the upper bound on the tidal deformability, in order to put a stringent constraint on the DM fraction for sub-GeV bosonic particles (see Fig. \ref{fraction}). We see that allowed region in which both the total maximum mass $M_{T_{max}}\geq2M_{\odot}$ and tidal deformability  $\Lambda_{1.4}\leq580$ constraints are satisfied is limited to relatively low DM fractions $F_{\chi}\lesssim5\%$ at the fixed value of the self-coupling constant $\lambda=\pi$.

The upper limit for the allowed DM fraction reduces significantly for light bosons going well below $1\%$. In our study we explore not only the conservative range of fractions achieved by accretion, but also alternative scenarios that predict large amount of DM inside a star. We showed that the existing data on compact stars do not contradict to the DM admixed NS scenario, and  every observed NS could potentially contain low  DM fraction  distributed in a core or in an extended halo. Moreover, DM admixed NS can serve as a satisfactory explanation for the unusual observational evidences on compact stars, e.g, the secondary object in the GW190814 event with the mass about $2.6M_{\odot}$ \cite{LIGOScientific:2020zkf}.

Note that for sub-GeV boson masses depending on the DM fraction, formation of both DM core and halo are possible for fixed $m_{\chi}$ and $\lambda$. In order to break the degeneracy and answer the question whether DM exists in the form of halo or core inside NSs, additional observable quantities other than the tidal deformability and the maximum mass are essential. The ongoing observations by the NICER  \cite{Raaijmakers:2021uju,Watts:2019lbs,Riley:2021pdl,Miller:2021qha} and LIGO/Virgo/KAGRA Collaboration \cite{LIGOScientific:2021qlt,LIGOScientific:2020zkf,KAGRA:2020tym,KAGRA:2019htd}, as well as the future Advanced Telescope for High Energy Astrophysics (ATHENA) \cite{Barcons:2012zb,Cassano:2018zwm}, the enhanced X-ray Timing and Polarimetry mission (eXTP) \cite{eXTP:2018kws,eXTP:2018anb,eXTP:2016rzs}, and the Spectroscopic
Time-Resolving Observatory for Broadband Energy X-rays (STROBE-X) \cite{STROBE-XScienceWorkingGroup:2019cyd,wilsonhodge2017strobex} 
telescopes may shed more light on the possible forms of bosonic SIDM in NSs.  

The outermost radius of a compact star with DM condensed in its core equals to the baryonic radius. However, observation of the outermost radius in the case of a DM halo formation imposes much bigger  challenges. Due to the fact that DM component distributed in the halo is undetectable through the spectroscopic measurements, the use of multimessenger astronomy is unavoidable. Thus, combining analysis of astrophysical and GW observations, as well as searches for microlensing or other gravitational effects close to the surface of compact stars, may give information about the halo structure around them.

Moreover, an additional piece of  valuable data is expected from the radio telescopes, e.g, the Karoo Array Telescope (MeerKAT) \cite{Bailes:2018azh}, the Square Kilometer Array (SKA) \cite{Watts:2014tja,Weltman:2018zrl}, and the Next Generation Very Large Array (ngVLA) \cite{2019clrp.2020...32D,Anu:2013}, that will look into the Galactic center. Despite a big dust extinction in the most central part of the Galaxy, we expect to find pulsars and magnetars in the region up to 70 pc from the center. This region may contain a high DM fraction, and therefore, can host DM admixed compact stars with  altered properties.

\vspace{0.5cm}
\section{acknowledgments}
S.S. and D.R. are very thankful to Fazlollah Hajkarim for insightful discussions and  comments on the draft. D.R. appreciates Sajad Khalili for his valuable assistance in writing the codes. V.S. acknowledges the support from the Funda\c c\~ao para a Ci\^encia e  a Tecnologia (FCT) within the projects No. UID/FIS/04564/2019, No. UID/04564/2020 and PHAROS COST Action CA16214. O.I. acknowledges the support from the Polish National Science Centre (NCN) under grant No. 2019/33/B/ST9/03059. It is part of a project that has received funding from the European Union's Horizon 2020 research and innovation program under grant agreement STRONG – 2020 - No 824093.
\section{appendix}
\label{sec-app}
\setcounter{section}{1}
\renewcommand{\thesection}{\Alph{section}}
\setcounter{equation}{0}
\renewcommand{\theequation}{\thesection.\arabic{equation}}
The DM EoS (\ref{e1}) has been originally obtained by \citet{Colpi:1986ye}. For the readers convenience we present its derivation in a flat space-time. Such a treatment is justified by the fact that gradients of metrics are small compared to the spatial scales. This can be shown by considering the gradient of the time-time component $g_{tt}$. Using the explicit expression from Ref. \cite{Oppenheimer:1939ne} one gets
\begin{eqnarray}
\frac{\partial g_{tt}}{\partial r}=-\frac{2g_{tt}}{P+\rho}\frac{dP}{dr}
\end{eqnarray}
with $P$ and $\rho$ being local pressure and energy density. We consider this quantity on the stellar surface where space-time is the most curved. In this case $g_{tt}=1-\frac{2M_{tot}}{R}$, $P=0$ and gradient of the pressure can be expressed using the TOV equation as $\frac{dP}{dr}=-\frac{\rho M_{tot}}{R^2g_{tt}}$. This yields $\frac{\partial g_{tt}}{\partial r}=\frac{2M_{tot}}{R^2}<\frac{1}{R}$, where the Schwarzschild limit $\frac{2M_{tot}}{R}=1$ was used on the second step. Thus, for a NS of radius $R\sim10~km$ within the spherical layer of thickness $\Delta r\sim1~m$, which is enough to be treated as a macroscopical scale allowing thermodynamic treatment of DM, relative deviation of the metrics from the flat one can be estimated as $\Delta r\frac{d g_{tt}}{dr}\lesssim\frac{\Delta r}{R}\sim 10^{-4}$.

The model of bosonic SIDM used in this work corresponds to the Lagrangian 
\begin{eqnarray}
\mathcal{L}=\frac{1}{2}\partial_\mu\phi^*\partial^\mu\phi
-\frac{m_\chi^2}{2}\phi^*\phi-\frac{\lambda}{4}(\phi^*\phi)^2.
\end{eqnarray}
Under the mean-field approximation deviation of the field bilinear $\phi^*\phi$ from its expectation value $\langle\phi^*\phi\rangle$ is assumed to be small. Therefore interaction term in $\mathcal{L}$ can be expanded up to terms linear in $\phi^*\phi-\langle\phi^*\phi\rangle$. Thus, the linearized mean-field Lagrangian becomes
\begin{eqnarray}
\mathcal{L}_{MF}
=\frac{1}{2}\partial_\mu\phi^*\partial^\mu\phi-\frac{m_\chi^{*2}}{2}\phi^*\phi+\frac{\lambda}{4}\langle\phi^*\phi\rangle^2.
\end{eqnarray}
First two terms of this Lagrangian describe free quasiparticles with the effective mass  $m_\chi^{*2}=m_\chi^2+\lambda\langle\phi^*\phi\rangle$. At zero temperature they form BEC (see e.g, \citet{2006ftft.book.....K}). The corresponding contribution to the total pressure is $\zeta^2(\mu^2_\chi-m_\chi^{*2})$, where $\zeta$ represents the amplitude of zero mode and $\mu_\chi$ is bosonic chemical potential. The last term in the mean-field Lagrangian does not include any dynamical fields but only the average of their product. Consequently, this term contributes the total pressure just as a constant term. Thus
\begin{eqnarray}
P=\zeta^2(\mu_\chi^2-m_\chi^{*2})+\frac{\lambda}{4}\langle\phi^*\phi\rangle^2.
\end{eqnarray} 
The amplitude of zero bosonic mode $\zeta$ attains a value, which maximizes the total pressure, i.e, $\frac{\partial p}{\partial\zeta}=0$. Condensate $\langle\phi^*\phi\rangle$ also maximizes the pressure leading to the condition  $\frac{\partial p}{\partial\langle\phi^*\phi\rangle}=0$. Finally, number density of bosons can be defined using the thermodynamic identity $n_\chi=\frac{\partial P}{\partial\mu_\chi}$. This leads to
\begin{eqnarray}
\label{A4}
&&2\zeta(\mu_\chi^2-m_\chi^{*2})=0,\\
\label{A5}
&&\lambda\left(-\zeta^2+\frac{\langle\phi^*\phi\rangle}{2}\right)=0,\\
\label{A6}
&&n_{\chi}=2\zeta^2\mu_\chi.
\end{eqnarray}
Eqs. (\ref{A5}) and (\ref{A6}) immediately yield $\zeta^2=\frac{\langle\phi^*\phi\rangle}{2}$ and $n_\chi=\langle\phi^*\phi\rangle\mu_\chi$. Since $\zeta\neq0$ in the condensate, then Eq. (\ref{A4}) gives  $\mu_\chi^2=m_\chi^{*2}$ or equivalently $\mu^2_\chi=m_\chi^2+\lambda\langle\phi^*\phi\rangle$. This defines $\langle\phi^*\phi\rangle=\frac{\mu_\chi^2-m_\chi^2}{\lambda}$. Consequently, total pressure and DM particle number density become
\begin{eqnarray}
P&=&\frac{1}{4\lambda}\left(\mu_\chi^2-m_\chi^2\right)^2,\\
n_\chi&=&\frac{\mu_{\chi}}{\lambda}\left(\mu_\chi^2-m_\chi^2\right).
\end{eqnarray}
Energy density can be found using the thermodynamic identity
\begin{eqnarray}
\rho&=&\mu_\chi n_\chi-P=\frac{3}{4\lambda}\left(\mu_\chi^2-m_\chi^2\right)^2+
\frac{m_\chi^2}{\lambda}\left(\mu_\chi^2-m_\chi^2\right)\nonumber\\
&=&3P+2m_\chi^2\sqrt{\frac{P}{\lambda}},
\end{eqnarray}
where on the second step $\mu_\chi^2-m_\chi^2$ was expressed through the pressure. This is a quadratic equation with respect to $\sqrt{P}$ yielding to
\begin{equation}
 \sqrt{P}=
\frac{1}{3}\left[-\frac{m_\chi^2}{\sqrt{\lambda}}\pm
\sqrt{\left(\frac{m_\chi^2}{\sqrt{\lambda}}\right)^2
+3\rho}\right].   
\end{equation}
The sign $"+"$ should be taken in order to provide positiveness of the solution. This gives exactly the Eq. (\ref{e1}).

\bibliography{references}

\end{document}